\documentclass[twocolumn,twocolappendix]{openjournal}
\renewcommand{\baselinestretch}{1.075} 
\shortauthors{Mendoza et al.}
\graphicspath{{./}{figures/}}
\usepackage{macros}

\begin{document}

\title{Differentiable Forward Modeling for Efficient and Accurate Shear Inference}

\author{\href{https://orcid.org/0000-0002-6313-4597}{Ismael Mendoza}$^{\star}$}
\email[$^\star$ Corresponding author: ]{ismael@umd.edu}
\affiliation{Argonne National Laboratory, 9700 South Cass Avenue, Lemont, IL 60439, USA}
\affiliation{Department of Astronomy, University of Maryland, College Park, MD 20742}
\affiliation{Department of Physics, University of Michigan, Ann Arbor, MI 48109}

\author{\href{https://orcid.org/0000-0002-5068-7918}{Axel Guinot}}
\affiliation{Department of Physics, McWilliams Center for Cosmology and Astrophysics, Carnegie Mellon University, Pittsburgh, PA 15213, USA}

\author{\href{https://orcid.org/0000-0001-7774-2246}{Matthew R. Becker}}
\affiliation{Argonne National Laboratory, 9700 South Cass Avenue, Lemont, IL 60439, USA}

\author{\href{https://orcid.org/0000-0001-8868-0810}{Camille Avestruz}}
\affiliation{Department of Physics, University of Michigan, Ann Arbor, MI 48109} 
\affiliation{Leinweber Institute for Theoretical Physics, University of Michigan, Ann Arbor, MI 48109, USA}

\author{\href{https://orcid.org/0000-0002-1590-6927}{Jean-Eric Campagne}}
\affiliation{Université Paris-Saclay, CNRS/IN2P3, IJCLab, 91405 Orsay, France}

\author{\href{https://orcid.org/0000-0002-7599-966X}{Natalia Porqueres}}
\affiliation{Université Paris-Saclay, Université Paris Cité, CEA, CNRS, AIM, 91191, Gif-sur-Yvette, France}

\author{\href{https://orcid.org/0000-0002-8505-7094}{Michael Schneider}}
\affiliation{Lawrence Livermore National Laboratory, 7000 East Avenue, Livermore, CA 94550, USA}

\author{\href{https://orcid.org/0000-0001-5082-4380}{Eleni Tsaprazi}}
\affiliation{Imperial Centre for Inference and Cosmology (ICIC) \& Astrophysics group, Department of Physics, Imperial College, Blackett Laboratory, Prince Consort Road, London SW7 2AZ, UK}

\author{The LSST Dark Energy Science Collaboration}

\begin{abstract}
    Forthcoming Stage-IV dark energy optical surveys, such as LSST, have the ambitious goal of measuring cosmological parameters at sub-percent precision. Realizing their full scientific potential requires very precise measurement of the cosmic shear signal and control of corresponding systematics.
    In this work, we present a modern implementation of the Bayesian shear inference framework in Schneider et al. (2015), in the case that the PSF and sky background are known.
    This framework automatically propagates the pixel-noise measurement error from each galaxy into the final shear estimate, and thus requires no external calibration to handle noise bias. 
    As a first application of this new implementation, we infer the cosmic shear posterior from simulated images consisting of isolated exponential galaxies with semi-realistic noise levels.
    In this simplified scenario, we estimate the absolute multiplicative bias $|m|$ of our approach to be below $0.9 \times 10^{-3} \, [3\sigma]$ when the intrinsic distribution of galaxy properties is known, and below $1.3 \times 10^{-3}\, [3\sigma]$ when these distributions are inferred alongside shear.
    Additionally, we make progress towards the algorithm's computational feasibility in the context of modern wide-field surveys, where billions of galaxies must be processed, by leveraging differentiable forward models of galaxies, gradient-based samplers, and GPUs.
    Our final galaxy-fitting MCMC produces $300$ effective samples of galaxy properties in $0.45$ seconds per galaxy using a single A100 GPU. 
    In the future, we seek to generalize our algorithm to handle selection, detection, and model shear biases so it can be applied to real survey data.
\end{abstract}

\maketitle

\section{Introduction} \label{sec:intro}

The gravitational potential generated by structure in our universe distorts the light from background source galaxies, altering their observed shapes due to weak gravitational lensing (WL) -- an effect called \textit{cosmic shear}. 
Cosmic shear traces the distribution of dark matter and the nature of dark energy, making WL one of the most powerful cosmological probes \citep{huterer2002weak, kilbinger2015cosmology, Prat2025WL}. 
Cosmic shear measurements have been successfully used in several Stage-III optical cosmological surveys, including KiDS \citep{asgari2021kids, wright2025kids}, DES \citep{amon2022desy3_shear,secco2022desy3_shear}, and HSC \citep{li2023hsc_shear_2pcf,dalal2023hsc_shear_spectrum} to constrain dark energy cosmological parameters to about a percent level. 
Modern Stage-IV optical surveys such as the Vera C. Rubin Observatory Legacy Survey of Space and Time (LSST) \citep{lsst2019overview}, Euclid \citep{euclid2011}, and the Roman Space Telescope \citep{spergel2015roman} have the objective of constraining dark energy to a sub-percent, which could pave the way for beyond $\Lambda$CDM physics.
Precise and accurate measurements of cosmic shear are critical to maximize the scientific return of these Stage-IV surveys.

Shear measurement is challenging: cosmic shear is about a $2\%$ change in a galaxy's axis ratio on an average cosmological line of sight \citep{bernstein2014bayesian, mandelbaum2018lensing}. 
LSST, for instance, has the ambitious goal of performing shear measurement with statistical errors below $1$ part in $1000$ of this $2\%$ \citep{mandelbaum2018lensing}.
Several shear measurement algorithms have been proposed over the years, but only a few have shown the potential to meet this challenging requirement. 
Currently, the dominant paradigm calibrates the ensemble shear estimator directly on the survey images by simulating the effect of a known shear, so-called \textit{self-calibration}.
This is the motivation for the \metacal \citep{huff2017metacal,sheldon2017metacal} algorithm, and its extension \metadetect \citep{sheldon2020metadetect, hoekstra2021metadetect, hoekstra2021accounting, sheldon2023metadetect_lsst}.
Another related approach is \anacal \citep{li2024diff, li2025noise}, which analytically calibrates the shear response using auto-differentiation and a suitable basis of FPFS shapelets \citep{li2022fsps}. 
Both \metadetect and \anacal have been shown to meet the LSST shear bias requirements, and \metacal and \metadetect have already been successfully applied to real data in DES \citep{desy1_shapes, gatti2021desy3_shear, yamamoto2025des_y6_shear}.

Despite the successes of self-calibration methods, they will face new challenges in the context of Stage-IV surveys.
For example, the coupled effects of blending in both shear and redshift calibration were shown in DES weak lensing analyses to be an important contribution to the mean multiplicative bias \citep{maccrann2022desy3}. This effect is likely to worsen with the increased depth of surveys like LSST \citep{sanchez2021effects, melchior2021challenge}.
Another set of systematics gaining relevancy are \textit{chromatic biases} \citep{meyers2015chromatic,eriksen2018wavelength,carlsten2018wavelength,kamath2019spectral,berlfein2025chromatic}. 
These arise because the Point Spread Function (PSF), a wavelength-dependent quantity, is modeled using stars which typically have different SEDs than galaxies. 
Currently, there is no \metacal extension accounting for these effects; \anacal cannot account for the aforementioned blending effect, but PSF-level corrections were recently implemented for chromatic biases and tested on Roman simulations \citep{berlfein2025chromatic}. 
It is likely that these methods will require external calibration from simulations as demonstrated in \cite{maccrann2022desy3}.  
Separately, both \metacal and \anacal require injecting additional noise to images as part of their procedure, significantly reducing the signal-to-noise ratio of measurements, and increasing systematic uncertainty. However, extensions of these methods that use deep field images are in development and might reduce the impact of this effect \citep{zhang2023deepmetacal, park2026deep}.
Given some of the current limitations in self-calibration methods, exploring other shear inference methods might still be fruitful.

Another set of algorithms for shear measurement use a forward model of galaxy images that includes relevant systematics. 
In principle, this approach reduces the bias on the shear estimator at the expense of increased model complexity and estimated uncertainty. 
A recent implementation of this idea is \lensmc \citep{euclid2024lensmc}, which samples galaxy ellipticities with an optimized MCMC and aggregates them across multiple galaxies to produce a final shear measurement. 
This algorithm has been tested on Euclid-like simulations and achieved reasonable shear measurement accuracy.
One downside of \lensmc is that the prescription to aggregate shapes into shear is an ad-hoc weighting and average scheme, and does not utilize the entire set of galaxy properties.
Another related method is \texttt{JIF} \citep{buchanan2023parametric} which used a physically-informed prior and selection corrections on the likelihood to obtain posterior samples of galaxy properties from images in the LSST DESC DC2 simulation \citep{abolfathi2021dc2}. 
This algorithm is statistically rigorous and established good probabilistic calibration for many of the galaxies in this simulation. However, the authors have not demonstrated that their samples can be used to estimate shear.
Both approaches might also be limited by their runtime in a Stage-IV survey: \lensmc was estimated to process one galaxy in $5$ seconds whereas \texttt{JIF} takes a few minutes. 

In this work we propose a new shear inference algorithm based on the Bayesian hierarchical framework from \cite{schneider2015hierarchical}. 
This framework infers shear from MCMC samples of galaxy properties obtained via model fitting based on a statistically rigorous formalism, and is parallelizable over disjointed sets of blended galaxy images.
Our implementation leverages differentiable forward models of galaxies in \jax \citep{jax2018github}, gradient-based samplers such as NUTS \citep{hoffman2014nuts}, and GPUs to speedup the MCMC galaxy property sampling.
Using this efficient implementation, we rigorously test our shear inference pipeline on a simple dataset of isolated parametric galaxy images with semi-realistic levels of shape and pixel noise, and measure its multiplicative bias using the shape noise cancellation method in \cite{pujol2019highly}.

In Section~\ref{sec:data} we describe the dataset of galaxy images simulated with \galsim \citep{galsim2015} that we use to test our shear measurement algorithm. Next, in Section~\ref{sec:methods} we describe our algorithm implementation in detail, laying out the statistical framework and the computational innovations. In Section~\ref{sec:results}, we present several tests we perform on our shear inference, including estimating its bias. Finally, we conclude and discuss future work in Section~\ref{sec:conclusions}.

\section{Dataset} \label{sec:data}

Our dataset for all our experiments consists of a set of $320$k isolated parametric galaxy images with size $63 \times 63$ pixels\footnote{This image size was chosen to be not too large, but still contain the vast majority ($\geq 99.5\%$) of the flux of all drawn light profiles.} and a pixel scale of $0.2''$, which is the same as LSST. All galaxy images are simulated with \galsim \citep{galsim2015}. Every simulated galaxy follows the exponential (disk) light profile, which corresponds to a S\'ersic profile with $n=1$ \citep{sersic1963influence}.
We choose the distributions of galaxy properties to obtain the semi-realistic SNR (signal-to-noise ratio) distribution presented below.

We choose parametric distributions for the flux $f$ and half-light-radius $s$ (in units of arcsecs) of our galaxies:
\begin{equation}
\begin{split}
    \log_{10} f &\sim \text{SkewNormal}(\mu_{f}=2.45, \sigma_{f}=0.4, a_{f}=14) \\
    \log_{10} s &\sim \mathcal{N}(\mu_{s}=-0.4, \sigma_{s}=0.05),
\end{split}
\label{eqn:lflux-lsize-dist}
\end{equation}
and ignore the correlation between the flux and size of galaxies in the context of this paper. The first two parameters of these distributions represent the mean and standard deviation, respectively. The log-flux has a skew normal distribution with shape parameter $a_{f} = 14$, which creates a smooth but rapid cutoff below approximately $\log_{10} f = 2.5$.
The PDF for the ``standard'' skew normal distribution for an arbitrary variable $x$ is:
\begin{equation}
    f(x;a) = 2 \phi(x) \Phi(ax),
\end{equation}
where $\phi$ is the PDF of the standard normal distribution, $\Phi$ is the CDF of the standard normal distribution, and $a$ is the shape parameter. To obtain the shifted and scaled version used for the log-flux above, the PDF is transformed as $f(x;a) \rightarrow f(z;a) / \sigma$ with $z \equiv (x - \mu) / \sigma$.

\begin{figure*}
    \centering
    \includegraphics[width=0.95 \textwidth]{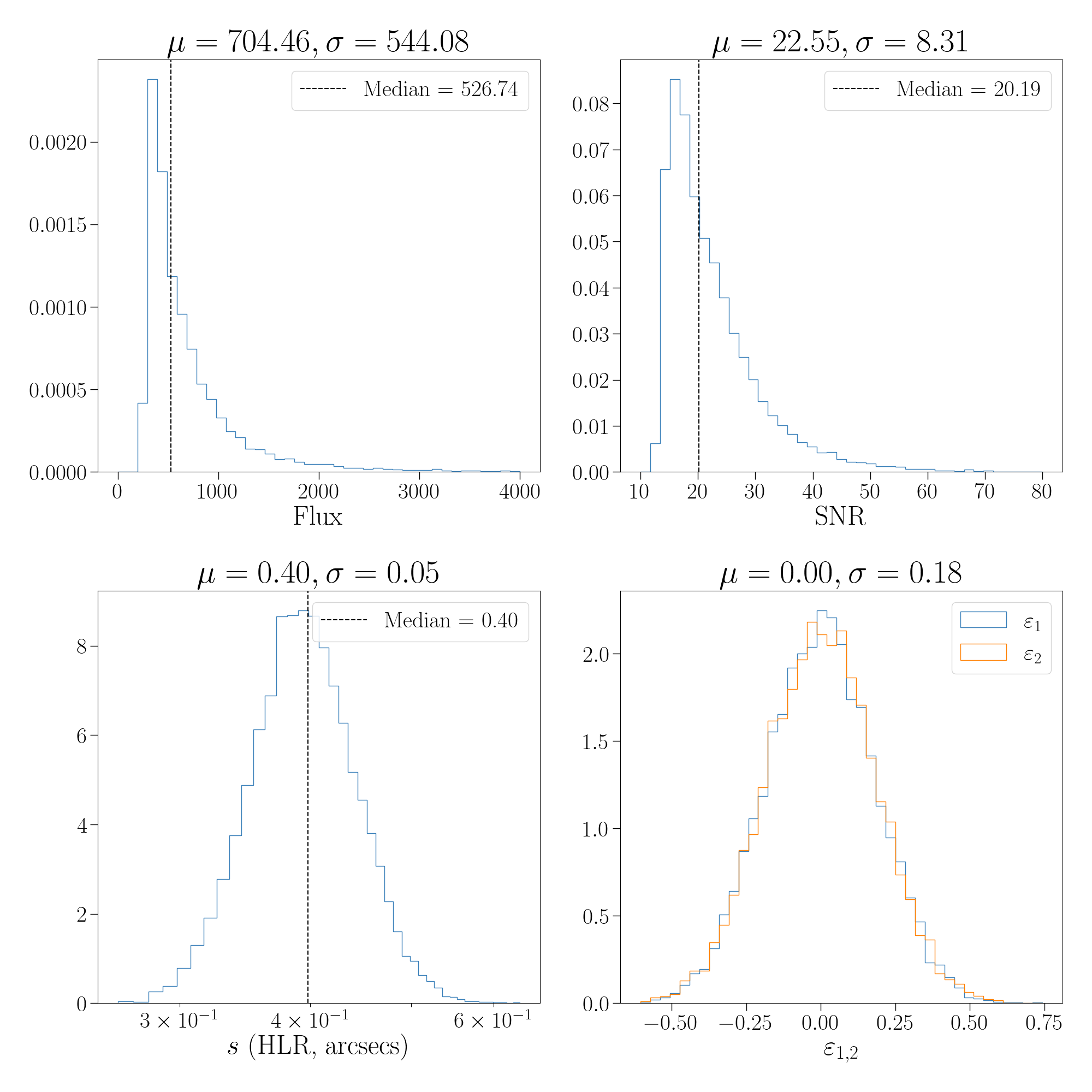}
    \caption{ 
        \textbf{Distribution of intrinsic galaxy properties.} In this figure we show distribution of intrinsic galaxy properties of our exponential galaxy dataset used in all our experiments. 
        The mean and sigma of each distribution are shown in the title of each distribution, and the median is shown as a dashed black line and in the legend. 
        For more details on this figure see Section~\ref{sec:data}.
    }
    \label{fig:gprops}
\end{figure*}

These choices for flux and size result in a SNR\footnote{The SNR is computed using the expression $\sqrt{\sum_{p} I_{p}^2 / b}$  where $I_{p}$ is the intensity at each pixel and $b$ is the sky level.} distribution with a median of about $20$.
We use a sky level value of $b = 1$ for all images and assume the sky-dominated approximation.\footnote{The actual values of flux and sky level used for our simulations are not physically meaningful and thus do not carry units. These were chosen to produce the SNR distribution displayed in Figure~\ref{fig:gprops}, which ultimately is what has an impact on the accuracy and precision of our measurements.} 
We have chosen this flux distribution to mimic observational effects that exclude very dim galaxies from the sample. The size distribution is too narrow compared to real data and was chosen for convenience. 
These result in an SNR distribution which is more optimistic than the forecasted weak-lensing sample in LSST, but includes galaxies with SNR down to $12$.

Galaxy ellipticities $(\eps_1, \eps_2)$ are sampled by first sampling the ellipticity magnitude $|\veps| \equiv \eps$ and orientation angle $\phi$ independently. Then, by taking the components of the complex ellipticity spinor:\footnote{We choose our ellipticity definition to match the `$\eps$' definition in \cite{bartelmann2001weak}, see Equation 4.10 in this paper. It also corresponds to the `reduced shear' $g$ definition in \galsim.}
\begin{equation}
    \veps = \eps_1 + i \eps_2 = \eps \cdot e^{2i\phi}.
\label{eqn:ellip-phi-transform}
\end{equation}
We choose the prior distribution on the magnitude of ellipticities $\eps \in (0, 1)$ to be the same distribution used in \cite{bernstein2014bayesian}, which has the form: 
\begin{equation}
    f_{e}(\eps) \propto \eps (1-\eps^2)^2 \exp(-\eps^2 / 2\sigma_{e}^2).
\label{eqn:ellip-mag-prior}
\end{equation}
This distribution corresponds to a truncated Gaussian with an additional factor of $(1-\eps^2)^2$ and support of $(0,1)$. This factor ensures that the prior has two continuous derivatives at the $\eps < 1$ boundary. 
We use a shape noise value of $\sigma_e = 0.2$. 
The orientation angle $\phi \in (0, \pi)$ is chosen to be uniformly distributed.

The centroid $(x, y)$ of each galaxy is uniformly distributed within the center pixel of each image. 
Finally, the PSF is the same for all images and it is modeled as a Gaussian with FHWM of $0.8$ arcsecs, which is close to the median value at zenith for LSST \citep{ivezic2019lsst}.
The distribution of intrinsic galaxy properties including flux, size, SNR, and ellipticities are shown in Figure~\ref{fig:gprops}. Their mean and standard deviation are shown in the corresponding subtitles, while the median is shown as dashed black line and the value is in the legend. 

In this paper we restrict to the case where both the PSF and background are known and constant. When applying this method in real survey data, separate survey pipelines could provide this information. 

\section{Methodology} \label{sec:methods}
We start by describing at a high-level the hierarchical Bayesian shear inference framework (Section~\ref{sec:hierarchy}). Next, we derive the shear likelihood which underpins our methodology in Section~\ref{sec:math}. Then, we detail our efficient implementation of this framework by taking advantage of differentiable forward modeling, gradient-based samplers, and GPUs (Section~\ref{sec:implementation}). Finally, we describe our method to measure multiplicative bias from inferred shear posteriors via the shape noise cancellation trick (Section~\ref{sec:shape-noise-cancellation}).

\subsection{Hierarchical Bayesian Shear Inference} \label{sec:hierarchy}
In this section we describe the algorithm for inferring the shear posterior $\prob{\vg | \bm{d}}$, where $\vg$ is the shear and $\bm{d}$ is a data vector corresponding to the pixels of a set of noisy individual galaxy images. 

Our procedure is a special case of the framework presented in \cite{schneider2015hierarchical} in the limit of isolated, individual galaxy images with a known PSF $\vpi$ and background sky-level $b$. It consists of two steps. 
First, we sample ellipticities and other galaxy properties from each individual galaxy image independently using a differentiable parametric model of galaxies and a gradient-based MCMC chain.
Second, we take these samples of galaxy properties and run an additional MCMC chain, which aggregates the samples across all the galaxies to evaluate the likelihood $\prob{\bm{d} | \vg}$ and obtain a posterior on cosmic shear.

Importantly, we allow for joint inference of the hyperparameters $\valpha$ describing the distribution of intrinsic galaxy properties along with the shear $\bm{g}$ using the same data $\bm{d}$. Thus, for full generality we include $\valpha$ in our target posterior: $\prob{\vg, \valpha | \vd}$. 
In our work, $\valpha$ includes the parameters characterizing the skew normal distribution of log-fluxes ($\mu_{f}$, $\sigma_{f}$, $a_{f}$), the normal distribution of log-sizes ($\mu_{s}$, $\sigma_{s}$), and the intrinsic ellipticity magnitude distribution ($\sigma_{e}$). 

\begin{figure}
    \centering
    \includegraphics[width=0.40 \textwidth]{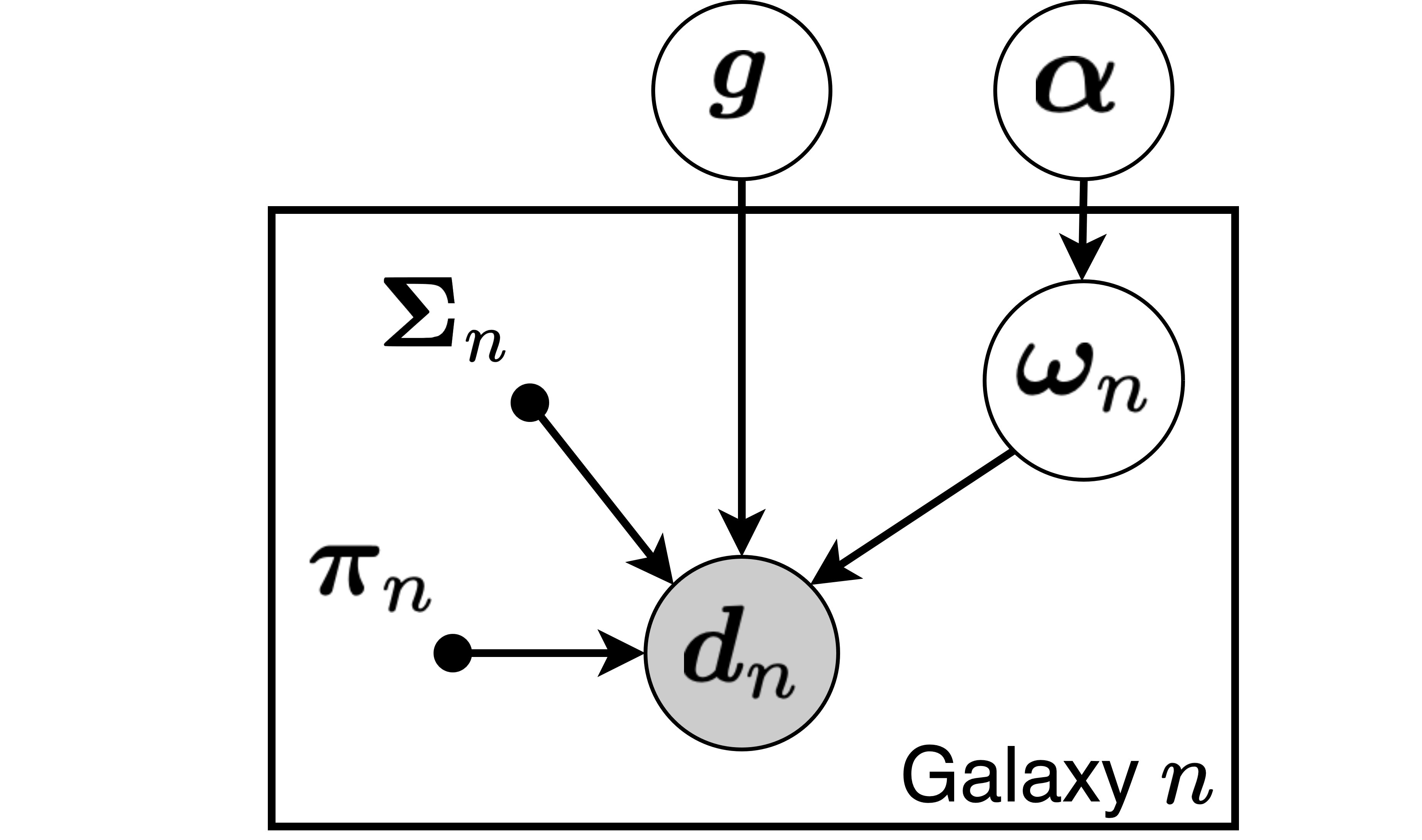}
    \caption{ \textbf{Probabilistic graphical model} representing our forward model for shear inference. A constant shear $\vg$ is applied to every galaxy with intrinsic parameters $\vomega_{n}$ with corresponding pixels $\vd_{n}$. The intrinsic properties of each galaxy are sampled from distributions characterized by hyperparameters $\valpha_{n}$, which includes the distribution of intrinsic ellipticities. The PSF $\vpi_{n}$ and noise covariance $\vSigma_{n}$ for each galaxy image are known and fixed.}
    \label{fig:pgm}
\end{figure}

Our model is visualized as a probabilistic graphical model in Figure~\ref{fig:pgm}. Each node represents a random variable, the dots represent fixed constants, arrows encode conditional dependencies, and boxes signal independent groups of variables. 
For generality, we include the possibility of different, known PSFs $\vpi_{n}$ and noise covariances $\vSigma_{n}$ for each galaxy, although these are both fixed in our experiments. Additionally, the noise is assume to be stationary and uncorrelated for all our experiments as described in the next section.

\subsection{Mathematical Formalism} \label{sec:math}

In this section, we derive a general version of the joint shear and intrinsic parameter likelihood $\prob{\vg, \valpha | \vd}$. 

First, let $\vomega_{n}$ be the vector of galaxy properties (flux, size, centroid, and shape) characterizing the $n$th galaxy in our sample, and $\vomega'_{n}$ be the sheared properties.\footnote{We do not model the effects of magnification and exclude it from our simulations, so that we can assume that cosmic shear $\vg$ only impacts galaxy ellipticity.} 
Specifically, our chosen ellipticity definition (Section~\ref{sec:data}) has the following response to shear \citep{seitz1997shear}:
\begin{equation}
    \bm{\eps}' \equiv h_{\bm{g}}(\veps) =  \frac{\bm{\eps} + \bm{g}}{1 + \bm{g}^{\star} \bm{\eps}},
\label{eqn:shear-transform}
\end{equation}
where $\bm{\eps}'$ is the sheared ellipticity, $\bm{\eps}$ is the intrinsic ellipticity of the galaxy, and $h_{\bm{g}}$ is the invertible `shear' transformation function. 
Now let $\bm{\phi}_{n}$ denote the vector of galaxy properties besides the shapes, ignoring magnification we have: 
\begin{equation}
\begin{split}
    \vomega_{n} &= (\vphi_{n}, \veps_{n}) \\
    \vomega_{n}' &= (\vphi_{n}', \veps_{n}') = (\vphi_{n}, h_{\vg}(\veps_n)).
\end{split}
\end{equation}
We can evaluate likelihood of an image $\vd_{n}$ given the corresponding properties $\vomega_{n}$ of the galaxy in that image. Given that most objects for lensing are background-dominated by a high-count sky level, we use the following Gaussian approximation to the Poisson distribution:
\begin{equation}
    \log \prob{\vd_{n} | \vomega_{n}, \vg} = -\frac{1}{2} \sum_{p=1}^{n_{\rm pixels}} \frac{(d_{np} - \bar{d}_{np} (\vomega_{n}, \vg, \vpi))^{2}}{b} + C,
\label{eqn:likelihood-pixels}
\end{equation}
where $\bar{\vd}$ is the noiseless parametric model for galaxies convolved with the PSF $\vpi$, and $C$ is a constant. In our chosen simulation setup the variance in every pixel is the same $b = 1$.\footnote{We currently ignore nearby pixel noise covariances from image coadding, unresolved background objects, or detector artifacts. In principle, these effects could be modeled and added to the likelihood. See Appendix~\ref{app:likelihood-extensions} for more details.}

In the context of this work, we require that shearing the galaxy models used to fit objects is equivalent to redrawing the models with the sheared parameters transformed as per Equation~\ref{eqn:shear-transform}. This is not true for all possible ways to parameterize models. Under this assumption, the likelihood can be rewritten as:
\begin{equation}
    \prob{\vd_{n} | \vomega_{n}, \vg} = \prob{\vd_{n} | \vomega'_{n}}.
\label{eqn:likelihood-requirement}
\end{equation}
Our goal is to write down a tractable expression for the likelihood $\prob{\bm{d} | \vg, \valpha}$, which will then allow us to run an MCMC to sample $\vg$ and $\valpha$. 
We start by introducing the \textit{sheared} galaxy properties by removing the marginalization: 
\begin{equation}
\begin{split}
    \prob{\vd \vert \vg, \valpha} & = \prod_{n=1}^{N_{\rm gal}} \prob{\bm{d}_{n} \vert \vg, \valpha}  \\
    & = \prod_{n=1}^{N_{\rm gal}} \int d \vomega'_{n} \prob{ \vomega'_{n}, \vd_{n} \vert \vg, \valpha},
\end{split}
\end{equation}
where the first line comes from the fact that the images are conditionally independent of each other given $\vg$ and $\valpha$ (see Figure~\ref{fig:pgm}). Next, we use the rules of conditional probability to obtain:
\begin{equation}
\begin{split}
    \prob{\vd \vert \vg, \valpha} & = \prod_{n=1}^{N_{\rm gal}} \int d \vomega'_{n} \prob{ \vomega'_{n} \vert \vg, \valpha} \prob{\vd_{n} \vert \vomega'_{n}, \vg, \valpha}  \\
    & = \prod_{n=1}^{N_{\rm gal}} \int d \vomega'_{n} \prob{ \vomega'_{n} \vert \vg, \valpha} \prob{\vd_{n} \vert \vomega'_{n}}.
\end{split}
\label{eqn:key-deriv-2}
\end{equation}
For the second line we note that conditioning on the sheared galaxy properties $\vomega'_{n}$ is redundant with conditioning alongside the shear $\vg$, and the galaxy properties' hyperparameters $\valpha$ are irrelevant once $\vomega'_{n}$ is known.
This expression for the likelihood remains impractical given the large dimensional integral over galaxy properties that would need to be performed at every step of the $\vg$ and $\valpha$ chain. 
We instead follow the approach in \cite{schneider2015hierarchical} and introduce the \textit{interim posterior distribution}:
\begin{equation}
    \iprob{\vomega'_{n} \vert \vd_{n}} \equiv \frac{1}{Z_{n}} \iprob{\vomega'_{n}} \prob{\bm{d}_{n} \vert \vomega'_{n}},
\label{eqn:interim-post}
\end{equation}
where $\iprob{\cdot}$ denotes the \textit{interim prior} and $Z_n$ is a normalization constant. 
The interim prior on measured galaxy properties can be chosen for computational convenience and does not need to be similar to the true prior. However, $\iprob{\cdot}$ must have the same domain as the true prior $\prob{\cdot \vert \valpha}$, and inference becomes more efficient the closer these two are. 

We then substitute $\prob{\bm{d}_{n} \vert \vomega'_{n}}$ from Equation~\ref{eqn:interim-post} into Equation~\ref{eqn:key-deriv-2} to obtain:
\begin{equation}
\prob{\vd \vert \vg, \valpha} = \prod_{n=1}^{N_{\rm gal}} Z_{n} \int d \vomega'_{n}  \frac{\prob{ \vomega'_{n} \vert \valpha, \bm{g}}}{\iprob{\vomega'_{n}}} \iprob{\vomega'_{n} \vert \vd_{n}}.
\end{equation}
Finally, we approximate the integral over $\vomega_{n}'$ of each galaxy with the usual Monte Carlo expression: 
\begin{equation}
    \prob{\vd \vert \vg, \valpha} \approx \prod_{n=1}^{N_{\rm gal}} \frac{Z_{n}}{K} \sum_{k=1}^{K} \frac{\prob{ \vomega'_{nk} \vert \valpha, \vg}}{\iprob{ \vomega'_{nk}}},
\label{eqn:likelihood-shear}
\end{equation}
where $\vomega'_{nk}$ are samples from the interim posterior $\iprob{\cdot \vert \vd_{n}}$ of the $n$th galaxy, and $K$ is the total number of samples from each galaxy. 
The numerator $\prob{ \vomega'_{nk} \vert \valpha, \vg}$ corresponds to the true prior of sheared galaxy properties and can be evaluated as follows: 
\begin{equation}
\begin{split}
    \prob{ \vomega'_{nk} \vert \valpha, \vg} &= \prob{ \vomega_{nk} \vert \valpha, \vg} \cdot \left\vert \frac{\partial \vomega_{nk}}{\partial \vomega'_{nk}} \right\vert  \\ &= \prob{ \vomega_{nk} \vert \valpha} \cdot \left \vert \frac{\partial \vomega_{nk}}{\partial \vomega'_{nk}} \right \vert  \\ 
    & = \prob{ \vomega_{nk} \vert \valpha} \cdot \left \vert \frac{\partial \veps_{nk}}{\partial \veps'_{nk}} \right \vert,
\end{split}
\label{eqn:true-prior-jacobian-shear}
\end{equation}
where in the first equality we changed variables in the probability density and picked up the absolute value of the determinant of the Jacobian $|\partial \vomega_{nk} / \partial \vomega'_{nk}|$. For the second equality we use the fact that the unsheared (intrinsic) galaxy properties samples, $\vomega_{nk}$, do not depend on the shear. 
Finally, the last equality uses the fact that only ellipticities transform when shear is applied. 
Note that the Jacobian can be evaluated through Equation~\ref{eqn:shear-transform}, and that it depends on the shear $\vg$.\footnote{We take advantage of the automatic differentiation system (Autodiff) in \jax to evaluate this Jacobian rather than analytically deriving an expression.}

The density $\prob{ \vomega_{nk} \vert \valpha}$ is the true prior on intrinsic galaxy properties. This prior could be learned from a subset of high-SNR galaxy observations (i.e., deep fields), and fixed when inferring shear on wide-field images. Another option is to jointly infer the prior with shear using the same set of wide-field images. 
In Section~\ref{sec:res:posteriors} we demonstrate that our framework can learn the parameters $\valpha$ of this distribution alongside shear using the same set of images.
Additionally, in Appendix~\ref{app:calibration}, we evaluate the calibration of the posteriors produced by this methodology in the context of a simplified ellipticity-only dataset.

\subsection{Phases and Settings} \label{sec:settings}

The likelihood in Equation~\ref{eqn:likelihood-shear} enables splitting shear inference in two phases: 
\begin{itemize}
    \item \textbf{Image Sampling Phase}: We sample the interim posterior (Equation~\ref{eqn:interim-post}) of each galaxy via MCMC independently to obtain lensed galaxy property samples $\{ \vomega'_{nk}\}$ using the interim prior $\mathcal{P}_{0}$ as the prior. The forward model used for sampling matches the exponential galaxy model used to create the dataset. 
    These samples only need to be obtained once under the interim prior and are fixed thereafter.

    \item \textbf{Shear Sampling Phase}: We run a MCMC chain to jointly infer the shear $\vg$ and the hyperparameters of the intrinsic galaxy property distributions $\valpha$ by evaluating the likelihood in Equation~\ref{eqn:likelihood-shear} using the samples from the image sampling phase: $\{\vomega'_{nk}\}$. 
\end{itemize}
We run four different `Settings' of these two phases on our dataset of $320k$ exponential galaxies:
\begin{itemize}
    \item \CaseA
    \item \CaseB
    \item \CaseC
    \item \CaseD
\end{itemize}
Here, the prefix `\texttt{shapes}' means that only galaxy ellipticities are sampled in the image sampling phase. On the other hand, `\texttt{all}' denotes that all galaxy properties (flux, shape, size, and centroid) are sampled in this phase. 
Next, the suffix `\texttt{fixed}' means only the shear is sampled while the galaxy prior hyperparameters $\valpha$ are fixed to their true values in the shear sampling phase. For instance, in \CaseA, the shape noise $\sigma_{e}$ is fixed to its true value used for generating the galaxy images ($\sigma_{e} = 0.2$), and used when evaluating the true prior on the shear likelihood in Equation~\ref{eqn:likelihood-shear}. Finally, `\texttt{free}' means that the relevant hyperparameters of the true galaxy prior are sampled alongside shear. 
We test all combinations of these choices for the image sampling phase and the shear sampling phase in Section~\ref{sec:results}. 
These settings are also summarized in Table~\ref{tab:settings}.

In the next section we discuss the implementation details of the MCMC chains in each phase. 

\begin{table*}
\centering
\begin{tabular}{lcccc}
\toprule
\textbf{Setting} 
& \multicolumn{2}{c}{\textbf{Image Sampling Phase}} 
& \multicolumn{2}{c}{\textbf{Shear Sampling Phase}} \\
\cmidrule(lr){2-3} \cmidrule(lr){4-5}
& \textbf{Sampled} & \textbf{Fixed} 
& \textbf{Sampled} & \textbf{Fixed} \\
\midrule
\CaseA & $\eps_{1}, \eps_{2}$ & $f, s, \Delta x, \Delta y$ & $g_{1}, g_{2}$ & $\sigma_{e}$ \\
\CaseB & $\eps_{1}, \eps_{2}$ & $f, s, \Delta x, \Delta y$ & $g_{1}, g_{2}, \sigma_{e}$ & - \\
\CaseC & $\eps_{1}, \eps_{2}, f, s, \Delta x, \Delta y$ & - & $g_{1}, g_{2}$ & $\sigma_{e}, a_{f}, \mu_{f}, \sigma_{f}, \mu_{s}, \sigma_{s}$ \\
    \CaseD & $\eps_{1}, \eps_{2}, f, s, \Delta x, \Delta y$ & - & $g_{1}, g_{2}, \sigma_{e}, a_{f}, \mu_{f}, \sigma_{f}, \mu_{s}, \sigma_{s}$ & - \\
\bottomrule
\end{tabular}
\caption{
    \textbf{Free and fixed parameters in each setting and phase.} 
    In this table we summarize the four shear inference settings tested in this work. 
    In the image sampling phase, galaxy properties are sampled by fitting individual galaxy images. In the shear sampling phase, shear is sampled alongside the parameters of the intrinsic galaxy property distributions. 
    Each setting has different properties that are sampled or fixed to their true value in the corresponding phases.
    The definition of every parameter can be found in Section~\ref{sec:data}. 
    For more detail on the settings see Section~\ref{sec:settings}.
}
\label{tab:settings}
\end{table*}

\subsection{Implementation} \label{sec:implementation}

Given the billions of galaxies that might need to be analyzed for weak lensing analysis in the era of Stage-IV surveys, our shear algorithms need to be highly efficient and scalable. 
The main bottleneck for our shear inference framework is the image sampling phase, as it requires one MCMC chain per galaxy in our dataset, which could easily become unfeasible with unoptimized chains.

We make our Bayesian framework efficient by leveraging differentiable forward modeling of galaxy images and GPU acceleration. 
Our forward model of galaxies is implemented in \jax \citep{jax2018github} and is provided by the python library \xgalsim.\footnote{This library is publicly available at \url{https://github.com/GalSim-developers/JAX-GalSim}, and was developed by the authors of this work.}
\xgalsim is a partial reimplementation in \jax of the popular weak-lensing simulation package \galsim \citep{galsim2015}. \xgalsim does not currently contain the full feature suite of \galsim, but available features have been tested using the unit tests available within \galsim, and are correct to about the same level of accuracy. 
Specifically, \xgalsim currently allows for rendering a core set of the parametric models available in \galsim and convolutions.
In our experiments, we have used \galsim to create our simulated exponential galaxy dataset and \xgalsim for the exponential galaxy forward model to evaluate the image likelihood (Equation~\ref{eqn:likelihood-pixels}).
The fact that our shear measurement algorithm outputs accurate shear measurements (Section~\ref{sec:results}) indicates that these two codes are largely numerically interchangeable in the regime we are using them. 

Using \xgalsim enables us to perform all image rendering operations in a GPU and parallelize over multiple galaxies simultaneously via \texttt{jax.vmap}. 
Furthermore, since our models are automatically differentiable, the likelihoods we use for MCMC sampling (Equations~\ref{eqn:likelihood-pixels} and~\ref{eqn:likelihood-shear}) are also automatically differentiable. Thus, we can take advantage of gradient-based MCMC like Hamiltonian Monte Carlo (HMC) \citep{neal2011mcmc} and the popular extension: No-U-Turn Sampler (NUTS) \citep{hoffman2014nuts}. 
These gradient-based approaches are known to exhibit higher efficiency and faster convergence for high dimensional distributions or those with complex geometries compared to classical random-walk MCMC \citep{betancourt2017mcmc}.
For all our experiments we use the \blackjax \citep{cabezas2024blackjax} implementation of the NUTS algorithm, which is also implemented in \jax and therefore enables our entire sampling procedure to run exclusively in GPUs. 

To be concrete, in the image sampling phase we run NUTS chains in parallel within the same GPU. Each chain runs $500$ warmup steps of dual averaging \citep{nesterov2009primal} to optimize the step size.\footnote{\blackjax attempts to follow the STAN implementation \citep{carpenter2017stan} of this algorithm as closely as possible.} The target distribution for the chains is the likelihood in Equation~\ref{eqn:likelihood-pixels} multiplied times the interim prior. 
We choose the interim prior $\iprob{\cdot}$ distribution to factorize over galaxy properties and be relatively uninformative: 
\begin{equation}
\begin{split}
    \log_{10} f &\sim U(-1, 9)\\
    \log_{10} s &\sim U(-2, 1)\\
    \Delta x &\sim \mathcal{N}(0, \sigma_{x} = 0.5) \\ 
    \Delta y &\sim \mathcal{N}(0, \sigma_{y} = 0.5) \\
    \eps &\sim p(\eps; \sigma_{e}=0.3) \\
    \phi &\sim U(0, \pi)
\end{split}
\label{eqn:interim-prior}
\end{equation}
where $U(\cdot, \cdot)$ represents a uniform distribution, and $\Delta x$ and $\Delta y$ are deviations from the true galaxy centroid.
Importantly, we sample $(\eps_1, \eps_2)$ rather than $(\eps, \phi)$ directly. The joint density is: 
\begin{equation}
\begin{split}
    \prob{\eps_1, \eps_2} &= \prob{\eps, \phi} \left \vert \frac{\partial (\eps, \phi)}{\partial (\eps_1, \eps_2)} \right \vert \propto f_{e}(\eps) / \eps \\
    &\propto (1-\eps^2)^2 \exp(-\eps^2 / 2\sigma_{e}^2)
\end{split}
\end{equation}
where $|\partial (\eps, \phi) / \partial (\eps_1, \eps_2)|$ is the determinant of the Jacobian determined by the transformation in Equation~\ref{eqn:ellip-phi-transform} and $f_{e}(\cdot)$ is the ellipticity magnitude prior density in Equation~\ref{eqn:ellip-mag-prior}.

We produce $300$ samples of galaxy properties following the warmup for all our experiments (samples produced during the warmup are discarded). 
We initialize the chains for each $n$ galaxy as follows: 
\begin{equation}
\begin{split}
    f_{n, \rm init} &= \sum_{p=1}^{n_{\rm pixels}} d_{np} \\ 
     \log_{10} s_{n, \rm init} &\sim U(\log_{10} s_{n} - 0.015, \log_{10} s_{n} + 0.015)\\
    \Delta x_{n, \rm init} &= 0 \\ 
    \Delta y_{n, \rm init} &= 0  \\
    \eps_{n, 1, \rm init} &\sim U(\eps_{n,1} - 0.1, \eps_{n,1} + 0.1)  \\
    \eps_{n, 2, \rm init} &\sim U(\eps_{n,2} - 0.1, \eps_{n,2} + 0.1)
\end{split}
\label{eqn:initialize}
\end{equation}
In other words, the flux is initialized as the sum of the flux in the image pixels, the size and shapes are initialized with small balls around the true value to represent some degree of measurement error, and the initial deviation from the true centroid is assumed to be $0$.\footnote{In general, chains initialized too far away from the truth will suffer from convergence issues. Based on our results in Section~\ref{sec:res:timing}, we find that our MCMC is robust to the level of deviation in Equation~\ref{eqn:initialize}. 
However, further testing will be needed in the context of real data as the true galaxy parameters are unknown.
We defer investigating more realistic initializations to future work, but one possibility is to find local maxima of the interim posterior using the methodology described in \cite{euclid2024lensmc}.}
The NUTS chains use a \mnd parameter of $5$ for all galaxies.\footnote{This parameter controls the maximum number of times the NUTS trajectory is doubled before returning if no U-Turn or divergence has occurred. See \citep{hoffman2014nuts} for more details.} 
For settings \CaseA and \CaseB, we only use the galaxy ellipticities to infer shear. In these cases, all of the other galaxy properties are fixed to their true values when running the MCMC chains and ellipticities are initialized in the same way as above. 

For the shear sampling phase, we use the likelihood in Equation~\ref{eqn:likelihood-shear} to sample the shear $\vg$ and the galaxy properties' hyperparameters $\valpha$.
The true prior in the numerator is defined by the data-generating process in Section~\ref{sec:data}. 
The interim prior in the denominator is defined in Equation~\ref{eqn:interim-prior}.  
It is critical to include the Jacobian factor for galaxy ellipticities in the numerator as shown in Equation~\ref{eqn:true-prior-jacobian-shear}.

For settings \CaseB and \CaseD, we use the following uninformative hyperpriors for the sampled $\valpha$ parameters: 
\begin{equation}
\begin{split}
    \sigma_{e} \sim U(10^{-4}, 1) \\ 
    \mu_{f} \sim U(-2, 6) \\
    \sigma_{f} \sim U(0, 2) \\ 
    a_{f} \sim U(-100, 100)\\
    \mu_{s} \sim U(-2, 2) \\ 
    \sigma_{s} \sim U(0, 0.5)
\end{split}
\end{equation}
where $\mu_{f}$, $\sigma_{f}$, and $a_{f}$ are the mean, sigma, and shape of the skew normal distribution for log-fluxes, and $\mu_s$ and $\sigma_{s}$ are the mean and sigma of the normal distribution of log-size (see Section~\ref{sec:data}). 
In all settings, for estimating shear $\vg$, we use a uniform prior on the unit disk. 
We note that the true prior on deviations from the true centroid $\Delta x$ and $\Delta y$ are the same as the interim prior by construction. Thus, their contributions to the likelihood cancel on a sample-by-sample basis and so we completely ignore the centroid samples and their priors in the shear sampling phase.

Finally, for the shear sampling phase we use a \mnd of $2$ for settings \CaseA and \CaseC. We use a value of $3$ for setting \CaseB and $7$ for \CaseD given their higher dimensionality. We keep $500$ warmup steps for all cases, except in \CaseD where we use $1000$ steps. 

\subsection{Multiplicative bias and shape noise cancellation} \label{sec:shape-noise-cancellation}

In the context of cosmological surveys, the requirement for the accuracy and precision of shear measurements is typically expressed in what is known as \textit{multiplicative} $m$ and \textit{additive} $c$ biases. Given a shear estimator $\hat{\vg}$ and the true shear applied $\vg$ these quantities are defined by the following equation:
\begin{equation}
    \hat{\vg} = (1 + m) \vg + c,
\end{equation}
which assumes that shears are weak enough that we can typically neglect higher-order terms. 
The requirement for Stage-IV dark energy surveys for constant-shear is $|m| < 2 \times 10^{-3}$ \citep{sheldon2023metadetect_lsst}.

We employ the shape noise cancellation trick introduced in \citep{pujol2019highly} to reduce the number of simulations needed to verify that our shear inference method fulfills this requirement. 
This trick consists of running two simulations with the exact same galaxy properties and pixel noise, but one simulation uses a small positive shear and the other uses the opposite shear. 
Next, by combining the shear estimator obtained from each simulation in a specific way, the multiplicative bias is estimated such that errors from shape noise and pixel noise are approximately canceled out. Reducing the impact of noise on our estimators reduces the total number of simulations that are needed.

Assuming that we apply a shear with only a non-zero $g_1$ component ($0.02$ for all our experiments), the multiplicative $m$ and additive $c$ biases can be estimated as follows: 
\begin{equation}
\begin{split}
    \hat{m} &= \frac{\hat{g}_{+, 1} - \hat{g}_{-, 1}}{2 | g |} - 1 \\
    \hat{c} &= \frac{\hat{g}_{+, 2} + \hat{g}_{-, 2}}{2}
\end{split}
\end{equation}
In our case, we choose our shear estimator to be the mean of the posterior $\prob{\vg | \vd}$,\footnote{We have verified that none of the conclusions of this work change if the median of the posterior is used as the shear estimator instead of the mean. In general, our resulting shear posteriors (see Figure~\ref{sec:res:posteriors}) visually seem very close to a Gaussian distribution. So we do not expect large differences in their mean, median, or mode.} and create two datasets $\vd_{+}$ and $\vd_{-}$ with the exact same $320$k galaxies in both sets of images, same noise realization, but with the opposite shear applied.

We estimate the error on both $\hat{m}$ and $\hat{c}$ using the bootstrap technique, where the shape noise cancellation procedure is applied to each galaxy resampling independently. 
For each resampling, the samples of galaxy properties from the original image sampling phase are reused to run the shear sampling phase. 
We use a total of $500$ bootstrap resampling for all cases with the exception of \CaseD due to computational limitations, where $200$ bootstraps are used. 
The error is then estimated by taking the standard deviation across the mean multiplicative bias of each bootstrap.
Our final estimate of the mean multiplicative bias uses the shear posterior from the original $320$k galaxies, not the resamplings.

\section{Results} \label{sec:results}

We present our results in this section. In Section~\ref{sec:res:timing} we evaluate individual galaxy chains on a subset of the full dataset. Specifically, we look at their convergence and sampling efficiency. 
Next, in Section~\ref{sec:res:posteriors} we present examples of shear and true prior's parameters posteriors obtained with our method. 
Finally, in Section~\ref{sec:res:multiplicative} we present the multiplicative bias of our method in the four different settings outlined in Section~\ref{sec:implementation}.

\subsection{Quality and Efficiency of Individual Galaxy Chains \label{sec:res:timing}}

We focus on evaluating the samples of galaxy properties in the image sampling phase with all galaxy properties free. These correspond to settings \CaseC and \CaseD (Section~\ref{sec:settings}). 
Every chain uses the likelihood given by Equation~\ref{eqn:likelihood-pixels}, the prior used is the interim prior from Equation~\ref{eqn:interim-prior}, and each chain is initialized independently using the procedure in Equation~\ref{eqn:initialize}. 
All NUTS chains in this section use $500$ warmup steps, $500$ sampling steps, and an \mnd of $5$. 

We first use a random subset of $10$k exponential galaxies from the full dataset to evaluate the convergence. For this purpose we use the $\hat{R}$ metric \citep{gelman1992inference}, which compares the intra- and inter-variance of the chains to assess convergence. According to the STAN manual, a value of $\hat{R}$ larger than $1.05$ suggests that the chains have not mixed well \citep{carpenter2017stan}. To obtain an $\hat{R}$ estimate for each galaxy we run four independent chains on the same image with different initializations (sampled independently from Equation~\ref{eqn:initialize}), and each using a different random seed. We use the \blackjax implementation of the $\hat{R}$ metric with the \texttt{potential\_scale\_reduction} function.
We find that $\hat{R}$ values for all galaxies in our subset is $< 1.01$.

\begin{figure}
    \centering
    \includegraphics[width=0.45 \textwidth]{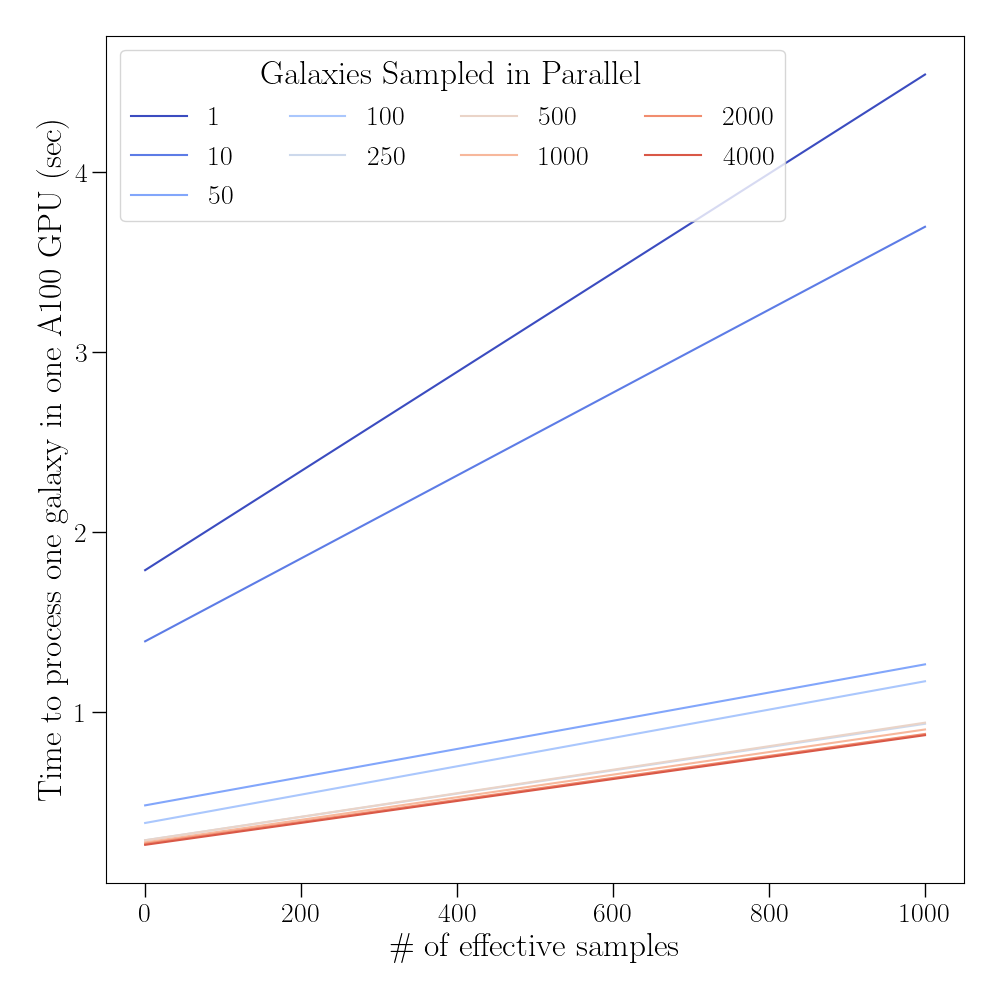}
    \caption{\textbf{Efficiency of sampling galaxy properties.} We show the time to obtain a specific number of effective samples \textit{per galaxy}, for a given number of galaxies (blue to red gradient in lines) being sampled in parallel in a single A100 GPU. 
    Producing zero effective samples corresponds to the time taken to warm up the galaxy chains, which takes a non-negligible amount of time. 
    See Section~\ref{sec:res:timing} for more discussion of this figure.}
    \label{fig:timing}
\end{figure}

Next we evaluate the efficiency of sampling galaxy properties by calculating the time taken to produced a given number of \textit{effective samples} in a A100 GPU.\footnote{This GPU has a total memory of about $40$GB, which allows us to run in parallel (with \texttt{jax.vmap}) over $5000$ individual galaxy NUTS chains.} 
We define the number of effective samples using the \textit{effective sample size} (or ESS) \citep[e.g.,][]{gelman1995bayesian}, which uses the autocorrelation length of the chain to estimate how many independent samples are produced. 
We summarize the results in Figure~\ref{fig:timing}, where we show the time taken per galaxy to obtain a specific number of effective samples, for a given number of galaxies being sampled in parallel on one A100 GPU. 
For a given number of chains (galaxies) $c$, each time curve is obtained by extrapolating the result of the original setup (which uses $500$ sampling steps) as follows:
\begin{equation}
    \frac{T_{\rm{warmup}}}{c} + \frac{T_{\rm{sampling}}}{c \cdot n_{\rm eff, 500}} \cdot n_{\rm eff}
\end{equation}
where $T_{\rm warmup}$ is the total time to warmup the $c$ chains in parallel, $T_{\rm sampling}$ is the total time taken to run $500$ sampling steps, $n_{\rm eff, 500}$ is the effective number of samples obtained after these $500$ steps (averaged over the six galaxy properties), and $n_{\rm eff}$ is the number of effective samples (the $x$-axis in Figure~\ref{fig:timing}).
For example, Figure~\ref{fig:timing} implies that running a single MCMC chain in the GPU to obtain $300$ total effective samples of galaxy properties takes approximately $2.6$ seconds. On the other hands, running $4000$ chains in the GPU simultaneously for the same number of effective samples takes a total time of $1784$ seconds, however, the time taken \textit{per galaxy} is lower than the former case: $0.45$ seconds. 

Overall, Figure~\ref{fig:timing} shows that the (effective) sampling efficiency increases as more chains are vectorized in the same GPU. However, this efficiency increases very slowly once more than $500$ chains are ran in parallel, even though the GPU has enough memory for additional chains.
This is surprising as chains are ran independently on the GPU, so naively we would expect that the efficiency keeps increasing until the GPU saturates. 
We hypothesize that this is caused by the NUTS sampler requiring a variable number of likelihood evaluations at each step, which means that all chains need to wait for the slowest chain at each iteration \citep{sountsov2024running}. This problem becomes more severe as the \mnd parameter is increased.

\subsection{Posteriors on Shear and True prior parameters } \label{sec:res:posteriors}

In this section, we show examples of the posteriors on shear $\vg$ and the intrinsic distribution hyperparameters $\valpha$ inferred by our method on our exponential galaxy dataset.

\begin{figure}
    \centering
    \includegraphics[width=0.45 \textwidth]{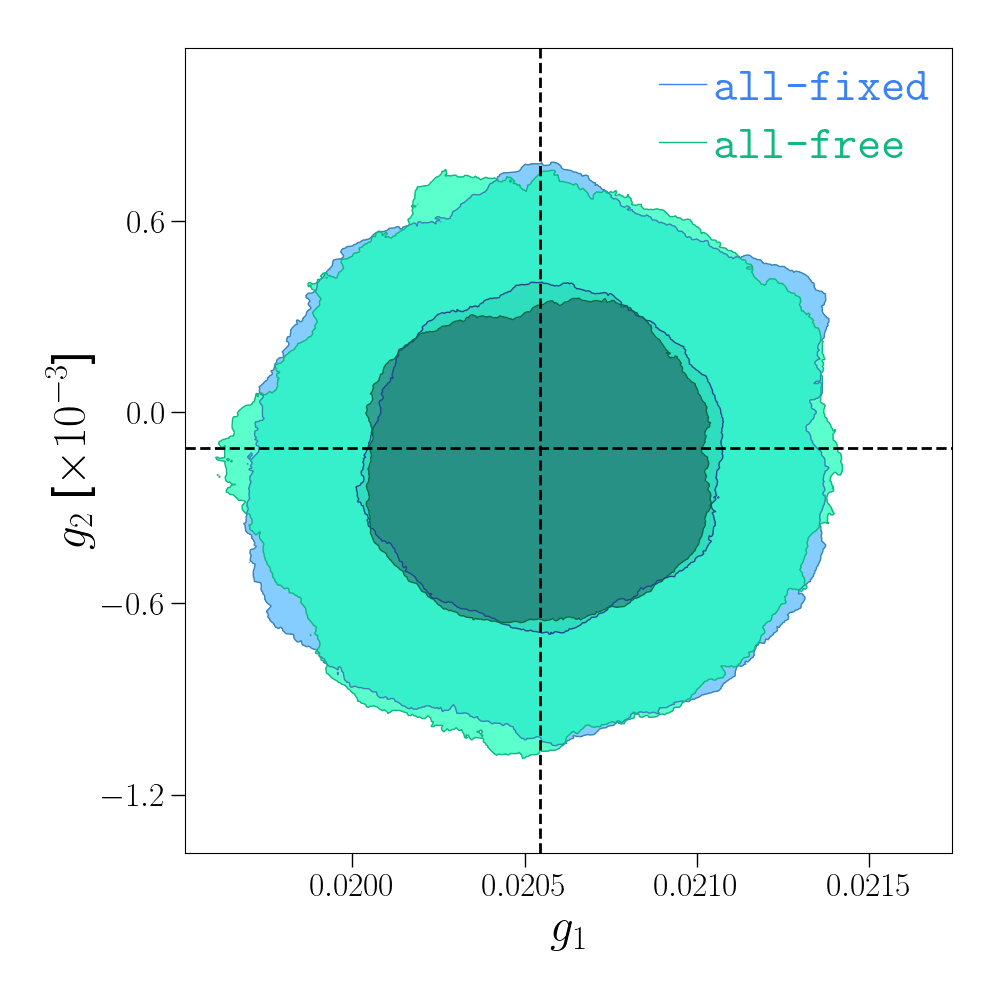}
    \caption{\textbf{Comparison of shear posteriors.} In this plot we compare the inferred shear posterior $1\sigma$ and $2\sigma$ contours obtained from our method applied to our dataset of $320$k isolated exponential galaxies with $g_1=0.02$ and $g_2 = 0$ true shear applied.
    The blue contours is the shear posterior inferred in setting \CaseC: the true prior of intrinsic galaxy properties is fixed and known, and only the shear is inferred.
    The green contours is the shear posterior in setting \CaseD: we jointly infer the shear alongside the hyperparameters of the intrinsic galaxy property distributions.
    The black dashed lines correspond to the true shear value plus the mean ellipticity of the dataset.
    See Section~\ref{sec:res:posteriors} for more details of this figure. 
    }
    \label{fig:shear-contours}
\end{figure}

First, in Figure~\ref{fig:shear-contours} we compare the shear posteriors inferred in settings \CaseC and \CaseD
applied to the same dataset of $320$k exponential galaxies with positive shear applied. In \CaseD, the shear is inferred alongside the intrinsic galaxy distribution hyperparameters $\valpha$. 
In both settings, we use the same samples of galaxy properties obtained from the image sampling phase to sample posteriors in the shear sampling phase. 
We see that the shear posteriors contours from each setting seem consistent with each other.
Both contours seem centered at the black dashed line plotted, which corresponds to the true shear value plus the mean ellipticity of the galaxy distribution. 
Importantly, we do not expect the contours to be centered at the true shear value, since they represent a single realization of the posterior inferred from a dataset with a finite amount of shape noise.
The width of the contours, however, accounts for shape noise so that the true shear value should be contained within them in most cases. 
For both of these posteriors, we find that the true shear value is within $2\sigma$ of their mean.

\begin{figure*}
    \centering
    \includegraphics[width=0.98\textwidth]{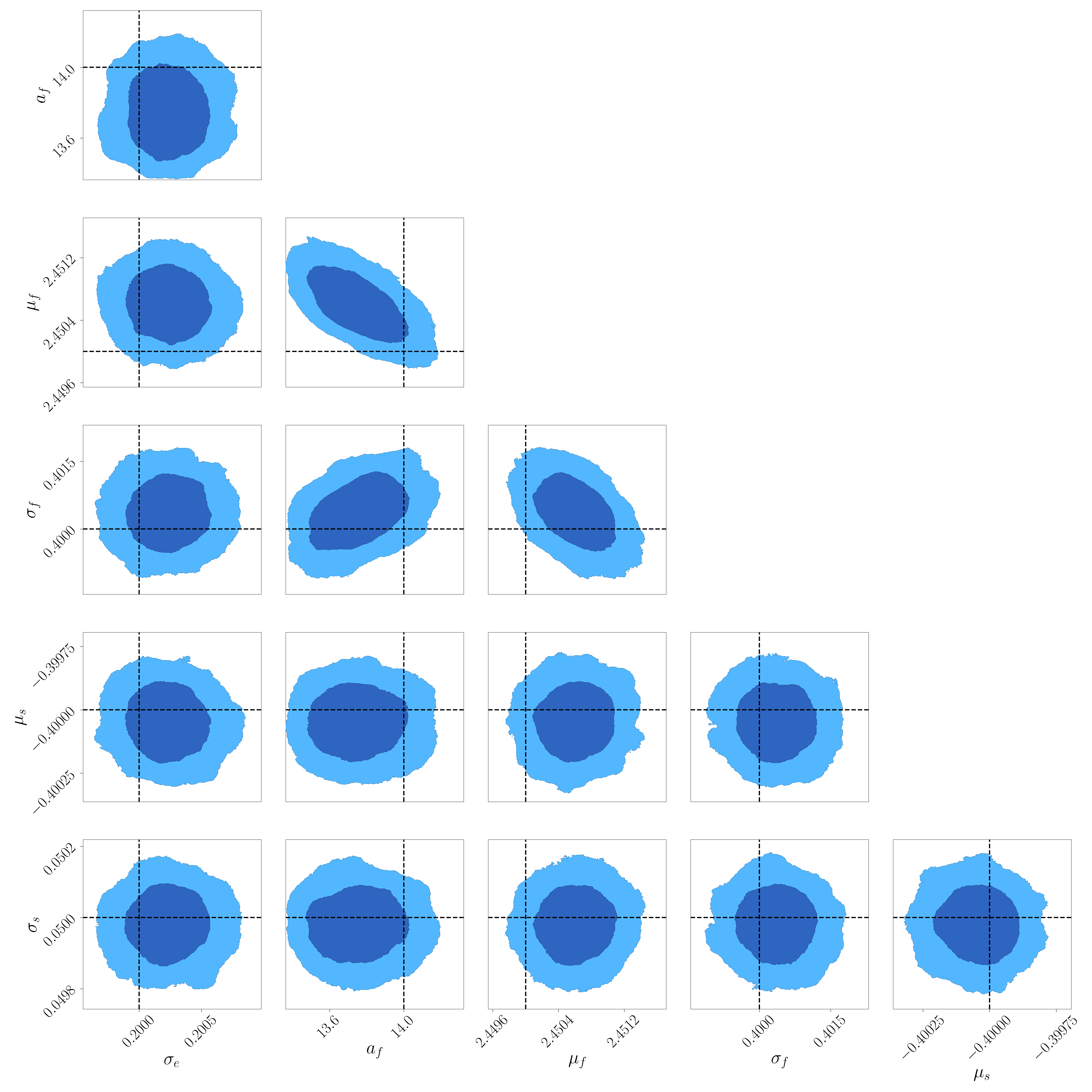}
    \caption{
        \textbf{Posteriors of the hyperparameters of distributions of intrinsic galaxy properties.} In this contour plot, we show $1\sigma$ and $2\sigma$ contours of the posterior of hyperparameters of galaxy properties' distributions based on $3000$ samples. Specifically, $\sigma_{\eps}$ corresponds to the scatter of the intrinsic ellipticity magnitude distribution (Equation~\ref{eqn:ellip-mag-prior}), $\mu_{f}$ and $\sigma_{f}$ are the mean and sigma of the lognormal distribution of fluxes, and $\mu_{\rm hlr}$ and $\sigma_{\rm hlr}$ are the mean and sigma of the lognormal distribution of half-light-radii (see Section~\ref{sec:data}). These hyperparameters are sampled jointly with shear using the same set of samples of galaxy properties. For more details see Section~\ref{sec:res:posteriors}.
    }
    \label{fig:hyper-contours}
\end{figure*}

Next, in Figure~\ref{fig:hyper-contours} we see the rest of the posterior contours inferred in \CaseD for intrinsic galaxy distribution hyperparameters $\valpha$. These are comprised of the skew normal parameters for the log-flux, the normal parameters for the log-size, and the shape noise $\sigma_{e}$.
For all variables we see that the true hyperparameters fall roughly within the $2\sigma$ contours of each posterior. All of the variables also seem uncorrelated, except for the parameters of the skew normal distribution on fluxes. 
We also see that most hyperparameters posteriors are very narrow. For instance, the scatter in the posterior of the shape noise parameter $\sigma_{e}$ is approximately $0.125\%$ of the true value of this parameter ($0.2$).\footnote{The reason that $\sigma_{e}$ is so precisely determined is that most of the galaxy shape data is driven by shape noise.} 
The fact that these parameters are so precisely determined explains why the width between posterior shear contours in settings \CaseC and \CaseD (Figure~\ref{fig:shear-contours}) is essentially the same.

Figures~\ref{fig:shear-contours} and \ref{fig:hyper-contours} demonstrate that we can successfully estimate shear and the intrinsic hyperparameters jointly on the same images without a significant increase in the error of the shear inferred. 
In the next section, we explore the multiplicative bias of our shear inference method in all settings.

\subsection{Multiplicative Bias} \label{sec:res:multiplicative}

In this section we present the multiplicative and additive bias estimated for our shear inference algorithm. These biases and their errors are estimated using the shape noise cancellation trick. Their errors are estimated with a bootstrap approach. See Section~\ref{sec:shape-noise-cancellation} for details. 

In Table~\ref{tab:bias} we show the results for each of our four settings. For each type of bias, the first number corresponds to the mean multiplicative bias and its error is the $3\sigma$ uncertainty obtained from the bootstrap procedure.
We see that all the multiplicative and additive biases are consistent with zero and within the Stage-IV weak lensing requirements.
The error on multiplicative bias seems to be larger for the cases when the true prior hyperparameters are free (cases \CaseB and \CaseD) compared to when these are fixed (cases \CaseA and \CaseC), as expected. 
The additive bias error seems to be approximately the same in all settings and consistent with zero.

\renewcommand{\arraystretch}{1.3}
\begin{table*}
    \centering
    \begin{tabular}{|c|c|c|c|c|c|}
        \hline
        \textbf{Experiment} & \textbf{Properties Used} & \textbf{Prior Known?} & \textbf{Multiplicative Bias $m / 10^{-3}$} & \textbf{Additive Bias $c / 10^{-3}$} \\
        \hline
        \texttt{shapes-fixed} & Ellipticities & Yes & $-0.361 \pm 0.698$ & $-0.176 \pm 0.969$ \\
        \texttt{shapes-free} & Ellipticities & No & $-0.267 \pm 0.913$ & $-0.172 \pm 1.079$ \\
        \texttt{all-fixed} & All & Yes & $-0.177 \pm 0.675$ & $-0.155 \pm 0.972$ \\
        \texttt{all-free} & All & No & $-0.313 \pm 0.893$ & $-0.16 \pm 0.975$ \\
        \hline
    \end{tabular}
    \caption{
        \textbf{Multiplicative and additive shear bias for different settings.} In this table we present the main results of this work, the multiplicative and additive bias (in units of $10^{-3}$) of our shear inference method in four different settings. 
        In all cases the shear inference method is applied on the same dataset of $320$k galaxies generated as detailed in Section~\ref{sec:data}.
        The mean multiplicative and additive bias is reported alongside with its $3\sigma$ error, which is estimated by via bootstrap resampling (Section~\ref{sec:shape-noise-cancellation}).
        Each setting samples different sets of galaxy properties (\texttt{shapes} or \texttt{all}), and either fixes or samples the hyperparameters of the intrinsic distributions of galaxy properties (\texttt{fixed} or \texttt{free}). See Section~\ref{sec:settings} and Table~\ref{tab:settings} for more details on these settings.
        See Section~\ref{sec:res:multiplicative} for additional discussion on this figure.
    }
    \label{tab:bias}
\end{table*}

\section{Summary and Discussion}\label{sec:conclusions}

\subsection{Key Results}

In this work, we have provided a modern re-implementation of the shear inference framework in \citep{schneider2015hierarchical} in the case that the PSF and sky background are known. 
We have rigorously tested our approach in a dataset of parametric, isolated galaxies with semi-realistic noise levels, and measured its multiplicative bias using the shape noise cancellation procedure in \cite{pujol2019highly}. 
We find that our algorithm's multiplicative bias is below $|m| < 2\times10^{-3}$, i.e., the threshold set as the LSST constant-shear requirement \citep{sheldon2023metadetect_lsst}. This is a promising result for our new method, but we emphasize that substantial development is still required before it can be applied to LSST real data (Section~\ref{sec:future-work}).
In addition, our approach leverages GPUs, differentiable forward modeling, and gradient-based samplers to improve the sampling efficiency of individual galaxy properties.

The main contributions of this work are as follows: 
\begin{itemize}
    \item We present an alternative derivation of the shear inference formalism from \cite{schneider2015hierarchical} in the special case of known PSF and background (Section~\ref{sec:math}).

    \item We present an efficient method to sample galaxy properties using model-fitting by taking advantage of a gradient-based sampler (NUTS), GPUs, and a differentiable model of parametric galaxies enabled by \xgalsim. 
    Our implementation produces one effective sample per galaxy in $0.6$ milliseconds after a burn-in phase that takes $0.26$ seconds per galaxy using a single GPU (Section~\ref{sec:res:timing}).

    \item We perform rigorous testing of our shear inference algorithm and use the shape noise cancellation trick \citep{pujol2019highly} to estimate its multiplicative bias. 
    We use a simple simulation consisting of $320$k isolated parametric galaxies, with their properties drawn from parametric distributions (Figure~\ref{fig:gprops}).
    If the true prior on intrinsic galaxy properties is known, we find a multiplicative bias $m$ of $(-0.177 \pm 0.675) \times 10^{-3}$ ($3\sigma$ confidence). 
    If the form of the prior is known, but its parameters are unknown, we find $m = (-0.313 \pm 0.893) \times 10^{-3}$ ($3\sigma$ confidence). 
    In both cases $m$ is consistent with zero and below the stated LSST requirement: $|m| < 2 \times 10^{-3}$. Our additive biases are also consistent with zero (Section~\ref{sec:res:multiplicative}).

    \item We demonstrate that our method accurately infers the hyperparameters of simple galaxy property distributions alongside with shear without incurring on multiplicative bias (Section~\ref{sec:res:posteriors} and ~\ref{sec:res:multiplicative}). 
    This is an important step towards the applicability of the method in real data, where the distribution of galaxy properties is not known.
\end{itemize}

\subsection{Other Bayesian Shear Measurement Methods}

Other proposed Bayesian shear measurement approaches include Bayesian Fourier Domain (\bfd) \citep{bernstein2014bayesian, bernstein2016accurate} and \lensmc \citep{euclid2024lensmc}.

The Bayesian Fourier Domain (BFD) method \citep{bernstein2014bayesian, sheldon2014practical, bernstein2016accurate} compresses images into a vector of moments $\bm{M}$ measured in Fourier space, and derives an analytical expression for the likelihood of these moments $P(\bm{M} \vert \bm{g})$ to ultimately arrive at the shear posterior. 
Our approach differs from \bfd in that we forward model galaxies in real space and use samples of the corresponding galaxy properties to perform shear inference. We require a model for the effect of shear on these galaxy properties, which in the context of this work is provided by Equation~\ref{eqn:shear-transform}. 
On the other hand, \bfd compresses the pixel information $\vd$ of galaxies into Fourier space moments $\bm{M}$, for which it is possible to analytically model the effect of shear. \bfd should be less sensitive to model bias than a model-fitting approach such as ours, since its fundamental `modeling' assumption is that the true distribution of galaxy moments is equal to the moment distribution of galaxies in a deep sub-survey.
However, because \bfd works in Fourier space, it currently cannot account for galaxy blends or pixel data that has been contaminated by cosmic rays or defects.
Our method can in theory include these effects by introducing them in our forward model.

\lensmc shares more similarities with the approach presented here, as \lensmc fits galaxy images and corresponding ellipticity samples are used to estimate shear. 
This method is an extension of an older Bayesian shear algorithm: \textsc{LENSFIT} \citep{miller2007fitting, kitching2008bayesian, miller2013bayesian}. 
\lensmc introduces a clever marginalization which reduces the dimensionality of the MCMC sampling and improves efficiency when fitting isolated galaxy images. Galaxy blends are handled separately by fitting the corresponding galaxies jointly.
However, the final shear estimator is based on a heuristic re-weighting of ellipticities, does not use all galaxy properties, and does not return a shear posterior. 
In contrast, as our method is based on \cite{schneider2015hierarchical}'s framework, it is statistically rigorous and returns a full shear posterior. 

\subsection{Advantages of Bayesian Shear Measurement}
\label{sec:advatanges}

The Bayesian forward modeling method presented here is fundamentally different from state-of-the-art self-calibration methods such as \metacal \citep{huff2017metacal} and its extension \metadetect \citep{sheldon2020metadetect}.
Whereas these methods compute a per-galaxy response that can be used to calibrate the shear bias incurred by any sensible shape measurement method at the population level, our algorithm infers the shear \textbf{directly} from a set of galaxy images by utilizing a suitable model of galaxies and a prior distribution of the intrinsic galaxy properties. 
Self-calibration approaches do not require galaxy distribution information since the response is computed via a \textit{perturbation} on measured galaxy shapes under an artificially applied shear.

In this work we have assumed that the true prior on galaxy properties is known or that it has a parametric form. In real surveys, this distribution will be more complex so another approach is needed. \cite{schneider2015hierarchical} suggests using a flexible distribution based on Dirichlet processes to learn the distribution of ellipticities and even accommodate different subsets of galaxies (e.g., blue vs. red galaxies) having different parameter distributions.  
Another flexible model for learning the prior might be normalizing flows (NFs) \citep{kobyzev2020normalizing, papamakarios2021normalizing}, which are based on neural networks and have been shown to learn even high-dimensional and multi-modal distributions. 
Regardless of the flexible model chosen to learn the intrinsic galaxy property distribution, there are two ways it can be employed to perform shear inference. 
First, we could learn the prior from a low-noise subset of the data, i.e., \textit{deep-fields}. This is the approach employed in \bfd \citep{bernstein2016accurate} and has the disadvantage that sample variance must be accounted for.
Another possibility is to learn the shear alongside the true prior distribution using the same subset of data, which becomes more difficult as the flexibility of the model distribution increases. 
This latter approach actually represents an opportunity for our methodology, as selection biases would (in principle) be automatically calibrated as part of this procedure. 
Previous work from \cite{congedo2025self} suggests that this type of ``Bayesian self-calibration'' could also apply more generally to \textit{any} type of multiplicative and additive bias. 
We plan to explore this idea further in future work.

More broadly, the primary advantage of our probabilistic framework is that it offers a clear recipe to address \textit{any} systematic biases in shear measurement: assert a flexible model for the relevant systematics and marginalize over the corresponding nuisance parameters \citep{schneider2015hierarchical}. 
As mentioned in the Introduction, there are at least two important systematics in the context of Stage-IV surveys that remain not fully accounted for by self-calibration methods: blending and chromatic effects.
For both of these, there are possible extensions to our methodology that could mitigate the corresponding shear bias.
In the case of blending \citep{maccrann2022desy3}, our hierarchical model could be augmented to include the tomographic bin assignment $\nu$ of each galaxy, with a corresponding shear $\vg_{\nu}$ for that bin. We will also need to generalize our model to use information from multiple bands to actually have tomograhic discriminatory power. This idea is similar to the one suggested in \cite{bernstein2016accurate}.
Regarding chromatic biases \citep{meyers2015chromatic}, one possibility is learning a wavelength dependent PSF model from nearby stars, and condition this model on the measured galaxy's colors. \cite{stone2025psf} demonstrated a Bayesian approach to inferring PSFs at the pixel-level, which could form a part of this procedure.

These and other generalizations to our method will likely be computationally complex and require significant optimization, but represent an opportunity to handle challenging types of shear biases in a statistically principled way.
This paper demonstrated how gradient-based samplers and GPUs can be leveraged to significantly reduce the computational cost of the classical implementation \citep{schneider2015hierarchical, mandelbaum2015great3}, and we expect that these tools will become increasingly essential as the dimensionality of the nuisance parameters increase. 

\subsection{Future Work and Current Limitations}
\label{sec:future-work}

Future work on this method will aim to enable its application on modern survey data. There are several significant challenges remaining.
One of them is to account for selection shear biases. These biases arise when the set of galaxies selected for shear estimation depends on the shapes of these galaxies \citep{kaiser2000shear, sheldon2017metacal, mandelbaum2018lensing}. In this case, the ellipticity distribution of galaxies in the sample will not follow the original distribution, since galaxy orientation is no longer uniformly distributed. Our current implementation cannot account for this type of bias since our prior distribution on ellipticities explicitly assumes that orientation angles are uniformly distributed (Section~\ref{sec:data}). 
An important type of selection bias is detection shear bias \citep{sheldon2020metadetect}. This bias occurs because the output of standard detection algorithms depends on the shear applied, which leads to a selection effect.
As mentioned in Section~\ref{sec:advatanges}, one potential way to handle selection bias is making the distribution of ellipticities more flexible to allow for skewness, and fitting the parameters of this distribution jointly with shear.

Another source of bias is \textit{blending} \citep{melchior2021challenge}, which refers to visually overlapping light sources in astronomical images. If all the sources in the galaxy blend are accurately detected, we could account for blends by jointly fitting the galaxies in the group during the image sampling phase. This is similar to the approach used in \lensmc.
Even if the centroid of every source is perfectly known, jointly fitting blended galaxies becomes more difficult, due to the correlated nature of their parameters and higher-dimensionality. It remains to be seen if our current sampling algorithms can efficiently handle this more difficult setting. 
Another complication in handling blended galaxies arises if these are too blended to be detected individually, so-called \textit{unrecognized blends} \citep{dawson2016ambiguous, sheldon2020metadetect}. 
In this case, our parametric model of galaxies cannot account for the resulting `hybrid source' and measured ellipticities will likely not transform under shear via Equation~\ref{eqn:shear-transform}, leading to biases.

Another important challenge to our method is the fact that there is no finite-parameter model that describes the light profiles of galaxies in the universe \citep{bernstein2014bayesian}. Thus, fitting survey data with a parametric model of galaxies can cause model bias \citep{bernstein2010shape, voigt2010limitations}.
We preliminarily investigate model bias in our setting by estimating the multiplicative bias when using a Spergel profile to fit the same $320$k exponential galaxy dataset. See Appendix~\ref{app:model} for details.
We find large and significant multiplicative biases when the (fixed) Spergel index $\nu$ is different from the value corresponding to an Exponential profile ($\nu=0.5$).
For instance, we see a bias of approximately $20\%$ when the Spergel index $\nu$ is close to $-0.6$, corresponding to a De Vaucouleurs profile, and of approximately $-5\%$ when $\nu$ is around $3$, which is closer to a Gaussian profile. These results further encourage future work to explore approaches to mitigate model biases. 
One possibility in the context of this study is to allow the Spergel index to vary alongside the other galaxy properties. 
For realistic galaxy images, adopting a flexible galaxy model learned from the data itself via machine learning \citep{lanusse2021deep, arcelin2021deblending, remy2022model, biswas2024madness} could be a path forward.
We also plan to use COSMOS Real galaxies in \galsim \citep{mandelbaum2012hst} to more realistically quantify and study model bias.

Finally, another avenue for development is further speeding up the galaxy property sampling. 
We have chosen to use No-U-Turn Sampler (NUTS) \citep{hoffman2014nuts} because it has been shown empirically to be successful in a diverse set of target distributions \citep{betancourt2017mcmc} and its \jax implementation is available in \blackjax \citep{cabezas2024blackjax}. 
However, given the adaptive nature of NUTS, different chains require a different number of likelihood evaluations. So that when running multiple chains in parallel in a single GPU, each chain will need to wait for all the others to complete a given step before advancing to the next. This makes NUTS a suboptimal sampling algorithm for GPUs \citep{sountsov2024running} as discussed in Section~\ref{sec:res:timing}. 
We plan to explore other recently developed approaches for gradient-based sampling designed to be GPU compatible to further improve the efficiency of our inference \citep{hoffman2021adaptive, hoffman2022tuning, onfroy2025mclmc}.

\section{Acknowledgments}

This paper has undergone internal review in the LSST Dark Energy Science Collaboration.
We are very grateful to our internal reviewers Xiangchong Li and François Lanusse. Additionally, we thank James Buchanan for helpful discussions and feedback on our results.

IM acknowledges support from DOE grant DE-SC009193. IM acknowledges the support of the Special Interest Group on High Performance Computing (SIGHPC) Computational and Data Science Fellowship. IM acknowledges support from the Michigan Institute for Computational Discovery and Engineering (MICDE) Graduate Fellowship. IM acknowledges support from the Leinweber Center for Theoretical Physics Summer Fellowship. IM acknowledges support from the U.S. Department of Energy, Office of Science, Office of Workforce Development for Teachers and Scientists, Office of Science Graduate Student Research (SCGSR) program. The SCGSR program is administered by the Oak Ridge Institute for Science and Education for the DOE under contract number DE-SC0014664. All opinions expressed in this paper are the authors' and do not necessarily reflect the policies and views of DOE, ORAU, or ORISE.
AG acknowledges the support of a grant from the Simons Foundation (Simons Investigator in Astrophysics, Award ID 620789).
JEC acknowledges use of GPU resources from French GENCI–IDRIS (Grant 2024-AD010413957R1).
ET acknowledges support by STFC through Imperial College Astrophysics Consolidated Grant ST/W000989/1.

The DESC acknowledges ongoing support from the Institut National de Physique Nucl\'eaire et de Physique des Particules in France; the Science \& Technology Facilities Council in the United Kingdom; and the Department of Energy and the LSST Discovery Alliance
in the United States.  DESC uses resources of the IN2P3 Computing Center (CC-IN2P3--Lyon/Villeurbanne - France) funded by the Centre National de la Recherche Scientifique; the National Energy Research Scientific Computing Center, a DOE Office of Science User Facility supported by the Office of Science of the U.S.\ Department of Energy under Contract No.\ DE-AC02-05CH11231; STFC DiRAC HPC Facilities, funded by UK BEIS National E-infrastructure capital grants; and the UK 
particle physics grid, supported by the GridPP Collaboration.  This work was performed in part under DOE Contract DE-AC02-76SF00515.

The submitted manuscript has been created by UChicago Argonne, LLC, Operator of Argonne National Laboratory (“Argonne”). Argonne, a U.S. Department of Energy Office of Science laboratory, is operated under Contract No. DE-AC02-06CH11357. The U.S. Government retains for itself, and others acting on its behalf, a paid-up nonexclusive, irrevocable worldwide license in said article to reproduce, prepare derivative works, distribute copies to the public, and perform publicly and display
publicly, by or on behalf of the Government. The Department of Energy will provide public access to these results of federally sponsored research in accordance with the DOE Public Access Plan. \url{http://energy.gov/downloads/doe-public-access-plan}

We also acknowledge the use of \texttt{numpy} \citep{numpy2020}, \texttt{scipy} \citep{scipy2020}, 
\texttt{matplotlib} \citep{matplotlib2007}, 
\jax \citep{jax2018github}, 
\blackjax \citep{cabezas2024blackjax}, 
ArviZ \citep{arviz2019}, 
\numpyro \citep{phan2019composable, bingham2019pyro}, \texttt{tqdm}\footnote{\url{https://github.com/tqdm/tqdm}}, 
Ruff\footnote{\url{https://github.com/astral-sh/ruff}},
Typer\footnote{\url{https://typer.tiangolo.com/}},
ChainConsumer \citep{Hinton2016},
Pytest \citep{pytest}, \texttt{conda-forge} \citep{conda_forge2015},
and \galsim \citep{galsim2015}.

\section*{Author Contributions}

\textbf{IM}: designed and executed experiments, developed \xgalsim, and wrote paper.
\textbf{AG}: designed experiments, advised and guided project, and reviewed and edited paper.
\textbf{MB}: advised and guided project, designed experiments, developed \xgalsim, and reviewed and edited paper.
\textbf{CA}: mentored and advised lead author. 
\textbf{JEC}: developed \xgalsim, evaluated several MCMC methodologies, and reviewed paper.
\textbf{NP}: statistics discussions, MCMC diagnostics, and paper review.
\textbf{MS} reviewed paper and provided feedback.
\textbf{ET}: statistical consultation, MCMC diagnostics, code validation, and paper review.

\section*{Data Availability}

The code to reproduce all of the data and plots presented in this work can be found in the following public GitHub (\url{https://github.com/LSSTDESC/BPD}) and in Zenodo (\url{https://zenodo.org/records/18225943}).
The repository for \xgalsim, which provides the differentiable model for parametric galaxies used in our results, is also publicly available in the following Github repository (\url{https://github.com/GalSim-developers/JAX-GalSim}) and in Zenodo (\url{https://zenodo.org/records/18939580}). 

\bibliography{references}
\bibliographystyle{aasjournal}

\appendix

\setcounter{equation}{0}
\renewcommand{\thesubsection}{\Alph{subsection}}
\renewcommand\theequation{A.\arabic{equation}}
\subsection{A: Posterior Calibration on Toy Dataset}
\label{app:calibration}

In this section, we characterize the calibration of the shear posteriors produced by our methodology using a simplified ellipticity-only dataset. 

Posterior calibration refers to the extent to which the errors predicted by an inference method are reliable. 
In the context of simulations where data $D \sim \prob{D \vert \theta}$ can be repeatedly generated from a prior sample $\theta \sim \prob{\theta}$, we can quantify the calibration of an inference method by calculating the fraction of times $q$ that $\theta$ falls within some probably density interval $I_{q'}$ of the inferred posterior $P(\theta \vert D)$. 
If posterior is well-calibrated, then $q$ (\textit{realized coverage}) and $q'$ (the mass contained by the interval; \textit{target coverage}) should equal each other for all $q'$. 
Coverage tests are related to other calibration tests found in the statistics literature \citep{geweke2004getting, cook2006validation, talts2018sbc, gelman2020workflow}.

To perform this test, we use a simplified ellipticity-only dataset, rather than the simulated image dataset (Section~\ref{sec:data}), for computational feasibility.
First, we sample ellipticities $(\eps_{1}, \eps_{2})$ using the same prior in Equation~\ref{eqn:ellip-mag-prior}. 
Next, these ellipticities are sheared using Equation~\ref{eqn:shear-transform}. 
Then, we transform this ellipticities to the $\eta$ parameterization \citep{bernstein2002shapes} with corresponding components:
\begin{equation}
    \eta_{1, 2} = 2\, \eps_{1,2} \tanh^{-1}(\eps) / \eps,
\label{eqn:eta-transform}
\end{equation}
where $\eps \equiv \vert \vec{\eps} \vert = \sqrt{\eps_{1}^2 + \eps_{2}^{2}}$. The advantage of using the $\eta$ parameterization is that the ellipticity domain is now the real numbers.
Thus, Gaussian noise can be added to each component independently to obtain noisy ellipticities:
\begin{equation}
    \tilde{\eta}_{1,2} \sim \mathcal{N}(\eta_{1,2}, \sigma_{\eta}),
\label{eqn:noisy-eta-like}
\end{equation}
where $\sigma_{\eta}$ is the independent scatter of each component representing a degree of measurement error.
We choose the value of $\sigma_{\eta} = 0.1$ by approximately matching the distribution of ellipticity samples obtained for the simulated image dataset.

Shear inference on these noisy ellipticity samples proceeds using the same general formalism detailed in Section~\ref{sec:math}. 
The shear sampling phase remains exactly the same, but the first phase is modified as we sample noiseless ellipticities $\eta_{1,2}'$ from noisy ellipticities $\tilde{\eta}_{1,2}$, rather than from noisy images.
Critically, the likelihood in the first phase needs to be adapted to reflect the procedure used to generate these noisy ellipticities.
The interim posterior (Equation~\ref{eqn:interim-post}) in this setting becomes:
\begin{equation}
\begin{split}
    \iprob{\eta_{1}', \eta_{2}' | \tilde{\eta}_{1}, \tilde{\eta}_{2}} &= \frac{1}{Z} \prob{\tilde{\eta}_{1}, \tilde{\eta}_{2} \vert \eta_{1}', \eta_{2}'} \iprob{\eta_{1}', \eta_{2}'} \\
    & = \frac{1}{Z} \prob{\tilde{\eta}_{1}, \tilde{\eta}_{2} \vert \eta_{1}', \eta_{2}'} \iprob{\eps_{1}', \eps_{2}'} | \frac{\partial (\eps_{1}', \eps_{2}')}{\partial (\eta_{1}', \eta'_{2})} |
\end{split}
\end{equation}
where $\eps_{1,2}'$ are the noiseless ellipticities in the $\eps$ parameterization. The first factor is the likelihood defined by Equation~\ref{eqn:noisy-eta-like}, the second factor is the interim prior on ellipticities which we choose to be the same as in Equation~\ref{eqn:interim-prior}, and the last factor is the absolute determinant of the Jacobian defined by the inverse of Equation~\ref{eqn:eta-transform}.

In terms of running the MCMC for each phase, we continue using the NUTS sampler but with a \mnd of $2$ for both phases, as the first phase is much simpler. We also initialize both phases at the true value for their corresponding parameter. Each noisy ellipticity is sampled $300$ times in the first phase.

\begin{figure*}[htbp!]
    \centering
    \includegraphics[width=0.95 \textwidth]{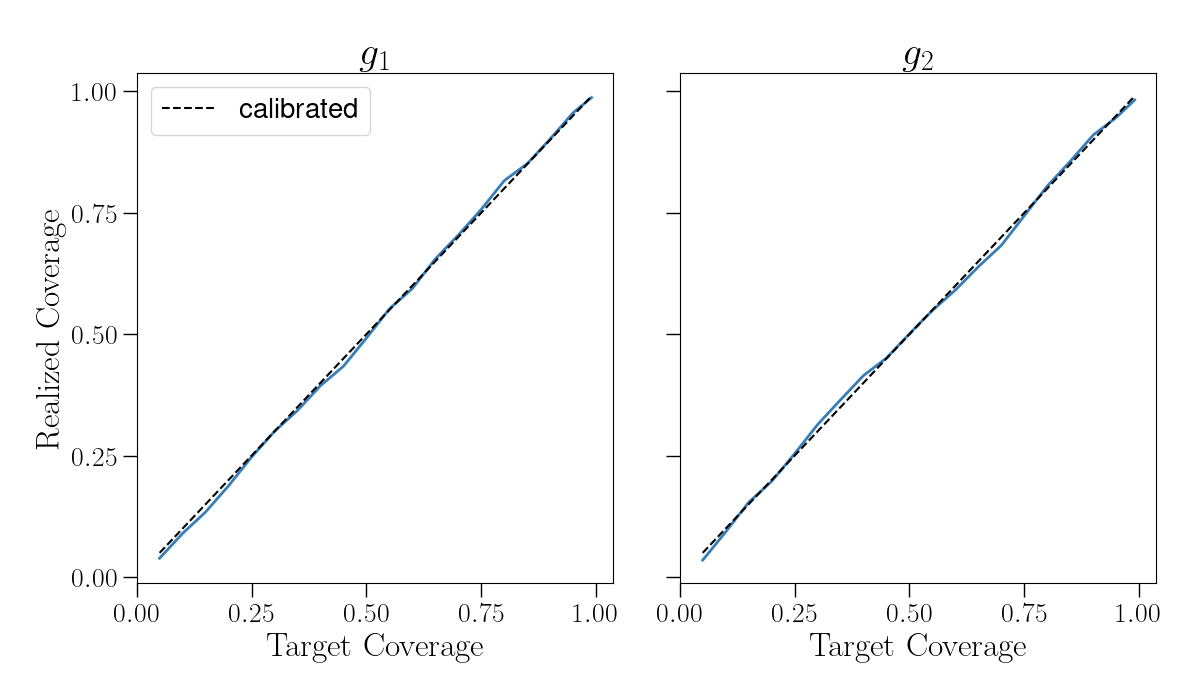}
    \caption{\textbf{Shear Posterior Calibration on Toy Ellipticities.} We present a coverage plot which we use to assess the calibration of the shear posteriors produced by our inference methodology (Section~\ref{sec:methods}) in the context of a simplified ellipticity-only dataset. 
    We first produce $1000$ shear posteriors by running our inference procedure on a $1000$ separate noisy ellipticity datasets each with a different shear applied. 
    For each of these posteriors and for a given target coverage, we create a probability density interval corresponding to this coverage. 
    The realized coverage is then the fraction of times that the true shear fell within this interval across all posteriors. We use $20$ equally spaced target coverages between $0.05$ and $0.99$ for this plot. 
    For more details on this Figure see Appendix~\ref{app:calibration}.
    \label{fig:calibration}
    }
\end{figure*}

We create the coverage plot in Figure~\ref{fig:calibration} by repeating the entire shear inference procedure $1000$ independent times on different randomly generated sets of $1000$ noisy ellipticities.
For each run we sample the true shear from a unit disk prior, and obtain $2000$ samples from the shear posterior.
Then, for a given target coverage $q'$, for each run, we use the \numpyro \texttt{hpdi} function to compute the highest probability density interval $I_{q'}$ from the posterior samples. 
Next, we compute the realized coverage $q$ as the fraction of times that each component of the true shear falls within $I_{q'}$. 
Finally, we plot $q$ and $q'$ in Figure~\ref{fig:calibration} using $20$ equally spaced target coverages between $0.05$ and $0.99$ as the blue curve.
The diagonal dashed line corresponds to a perfect calibration i.e., $x=y$. 
From this plot we see that the blue curve is really close to the dashed line, indicating that the posteriors produced by our inference method in this setting are well calibrated.
Overall this coverage experiment serves as a self-consistency test for our implementation and methodology. 

\subsection{B: Model Bias Study with Spergel Profiles}
\label{app:model}

We repeat the procedure to calculate multiplicative bias for the same $320$k exponential galaxies described in Section~\ref{sec:res:multiplicative} in the \CaseC case (with the same MCMC settings), but use a Spergel profile as the forward model with various (fixed) Spergel indices $\nu$. A Spergel index of $\nu=0.5$ corresponds exactly to an exponential profile, and a Spergel index of $\nu=-0.6$ to a De Vaucouleurs profile. We also compute the multiplicative bias for the case of a Gaussian profile forward model.

\begin{figure*}[htbp!]
    \centering
    \includegraphics[width=0.95 \textwidth]{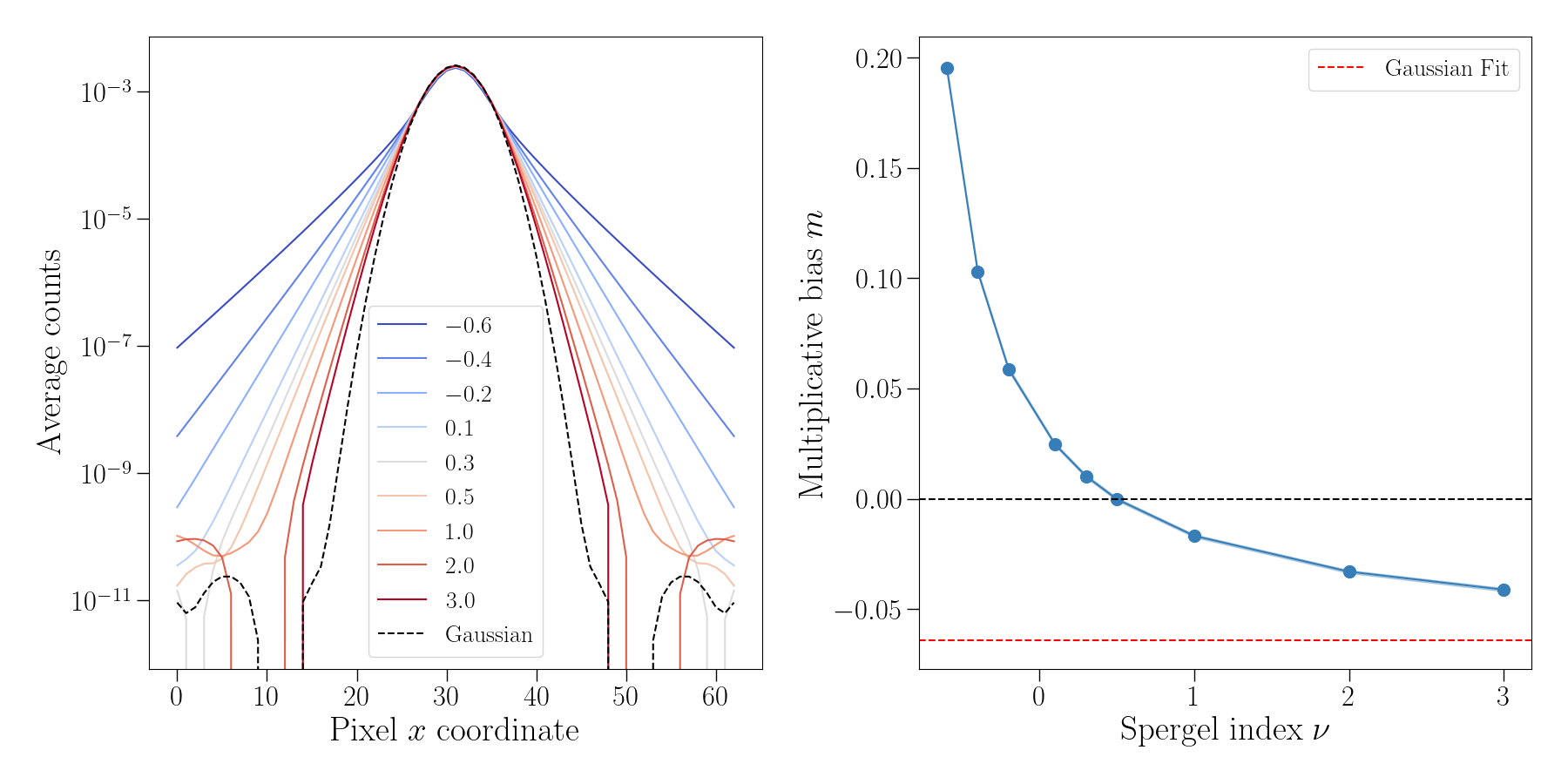}
    \caption{\textbf{Multiplicative bias as a function of Spergel index.} On the right of this figure, we plot the multiplicative bias $m$ calculated when using a Spergel profile with different (fixed) indices $\nu$ as the forward model in the \CaseC case, instead of the correct Exponential profile. 
    The red dashed line is multiplicative bias obtained when using a Gaussian profile forward model.
    On the left, we plot the (vertically) averaged circular profile for all the models used. All profiles plotted are noiseless, have a total flux of $1$, and a half-light-radius of $0.4$. 
    See Appendix~\ref{app:model} for more details on this figure.
    }
    \label{fig:model-bias}
\end{figure*}

The multiplicative biases as a function of Spergel index $\nu$ are displayed in Figure~\ref{fig:model-bias}. 
The red dashed line indicates the multiplicative bias for the Gaussian model. 
The shaded regions (barely visible) are the $3\sigma$ errors computed with a bootstrap procedure, but only using $100$ bootstraps due to computational limitations.
As a reference, on the left plot, we also include the average one-dimensional profile of each of the models we tested, where each of these have a total flux of $1$ and half-light-radius of $0.4$. 
From the right plot, we see that the magnitude of the multiplicative bias increases as we move away from the index corresponding to an Exponential profile $\nu=0.5$, as expected. 
The multiplicative bias at $\nu=0.5$ is consistent with the value in the third row of Table~\ref{tab:bias}.
As we decrease $\nu$ away from $\nu=0.5$ and the profile becomes closer to a De Vaucouleurs, the multiplicative bias increases rapidly reaching values of over $10\%$. 
On the other hand, as $\nu$ increases, the multiplicative bias values become large and negative dipping below $-3\%$, but at a slower rate than the other side. 
Finally, we see that the multiplicative bias for the Gaussian fit is the most negative out of all the models we tested.
Overall, we find significant and large multiplicative model biases for our algorithm when the incorrect model is used. In the context of varying the Spergel index, the multiplicative bias seems to increase steeply with deviations from the true Spergel index.

\subsection{C: Extensions to the Pixel Likelihood}
\label{app:likelihood-extensions}

The Gaussian likelihood used to fit the galaxy model in the image sampling phase (Equation~\ref{eqn:likelihood-pixels}) assumes that the noise is constant and uncorrelated across pixels. However, noise present in real world astronomical images could be spatially variable or correlated due to processes like image coaddition.
Fortunately, this Gaussian likelihood can be generalized to handle these effects by introducing a pixel noise covariance matrix $\vSigma$:
\begin{equation}
    \log \prob{\vd | \vomega, \vg} = -\frac{1}{2} (\vd - \bar{\vd})^{T} \vSigma^{-1} (\vd - \bar{\vd}) + C,
\label{eqn:covar-pixel-likelihood}
\end{equation}
where $\vd$ is the target exposure, $\bar{\vd} = \bar{\vd}(\vomega, \vg, \vpi)$ is the galaxy model drawn, and $C$ is a constant (we have dropped the galaxy subscript $n$ for simplicity).
In practice, $\vSigma$ is usually estimated as part of the image processing of the survey by, e.g., using measurements from the empty regions of the sky \citep{galsim2015}.

The pixel likelihood can also be extended to allow for jointly fitting multiple exposures which contain the same object. Given that the exposures are independent measurements, the likelihood will becomes a sum over their individual contributions:
\begin{equation}
    \log \prob{\vec{D}| \vomega, \vg} = -\frac{1}{2} \sum_{j} (\vd_{j} - \bar{\vd}_{j})^{T} \vSigma_{j}^{-1} (\vd_{j} - \bar{\vd}_{j}) + C,
\end{equation}
where $j$ indexes the exposures, $\vec{D} = \{d_{j}\}$ is the collection of target exposures, $\vSigma_{j}$ is the pixel covariance for exposure $j$, and $\bar{\vd}_{j} = \bar{\vd}_{j}(\vomega, \vg, \vec{\pi}_{j})$ is the model which uses the same galaxy parameters and shear across exposures, but different PSF for each $\vpi_{j}$. For this type of approach, it is common to first run the detection on the coadd combining the exposures, and then fit the rest of the galaxy properties while keeping this centroid fix using this likelihood.

For both of the extensions presented here, only the image sampling phase would need to be modified while the shear sampling phase would remain unchanged.
We plan to investigate these and other generalizations to the pixel likelihood in upcoming papers.

\end{document}